\documentclass{pasa}%

\usepackage{graphicx, longtable, lscape}

\title[The MAGPI Survey]{The MAGPI Survey - science goals, design, observing strategy, early results and theoretical framework}

\author[Foster, Mendel, Lagos, Wisnioski, Yuan et al.]{C. Foster,$^{1,2}$\thanks{E-mail: caroline.foster@sydney.edu.au}\, J. T. Mendel,$^{3,2}$ C. D. P. Lagos,$^{4,2,5}$ E. Wisnioski,$^{3,2}$ T. Yuan,$^{6,2}$ 
F. D'Eugenio,$^{7}$
T. M. Barone,$^{3,1,2}$ K. E. Harborne,$^{4,2}$ S. P. Vaughan,$^{1,2}$ F. Schulze,$^{8,9}$ R.-S. Remus,$^{8}$ A. Gupta,$^{10,2}$  
F. Collacchioni,$^{11,12}$ D. J. Khim,$^{13}$ P. Taylor,$^{3,2}$ 
R. Bassett,$^{6,2}$ S. M. Croom,$^{1,2}$ R. M. McDermid,$^{14,2}$
A. Poci,$^{15,14}$ 
A. J. Battisti,$^{3,2}$ 
J. Bland-Hawthorn,$^{1,2}$
S. Bellstedt,$^{4}$ M. Colless,$^{3,2}$ L. J. M. Davies,$^{4}$ 
C. Derkenne,$^{14,2}$
S. Driver,$^{4,2}$ A. Ferr\'e-Mateu,$^{16,6}$
D. B. Fisher,$^{6,2}$ E. Gjergo,$^{16}$ E. J. Johnston,$^{18,19}$
A. Khalid, $^{1}$ 
C. Kobayashi,$^{20,2}$ S. Oh,$^{3,2}$ Y. Peng,$^{21}$
A. S. G. Robotham,$^{4,2}$ P. Sharda,$^{3,2}$ S. M. Sweet,$^{22,2}$ E. N. Taylor,$^{6}$
K.-V. H. Tran,$^{10,2}$ J. W. Trayford,$^{23}$ J. van de Sande,$^{1,2}$ S. K. Yi,$^{13}$ 
L. Zanisi,$^{24}$
\affil{$^{1}$Sydney Institute for Astronomy, School of Physics, A28, The University of Sydney, NSW, 2006, Australia}
\affil{$^{2}$ARC Centre of Excellence for All Sky Astrophysics in 3 Dimensions (ASTRO 3D)}
\affil{$^{3}$Research School of Astronomy and Astrophysics, Australian National University, Canberra, ACT 2611, Australia}
\affil{$^{4}$International Centre for Radio Astronomy Research, The University of Western Australia, 35 Stirling Highway, Crawley, WA 6009, Australia}
\affil{$^{5}$Cosmic Dawn Center (DAWN).}
\affil{$^{6}$Centre for Astrophysics and Supercomputing, Swinburne University of Technology, PO Box 218, Hawthorn, VIC 3122}
\affil{$^{7}$Sterrenkundig Observatorium, Universiteit Gent, Krijgslaan 281 S9, B-9000 Gent, Belgium}
\affil{$^{8}$Universit\"ats-Sternwarte M\"unchen, Fakult\"at f\"ur Physik, LMU M\"unchen, Scheinerstr.\ 1, D-81679 M\"unchen, Germany}
\affil{$^{9}$Max Planck Institute for Extraterrestrial Physics, Giessenbachstra{\ss}e 1, D-85748 Garching, Germany}
\affil{$^{10}$School of Physics, University of New South Wales, Kensington, Australia}
\affil{$^{11}$Instituto de Astrof\'isica de La Plata (CCT La Plata, CONICET,UNLP), Observatorio Astron\'omico, Paseo del Bosque, B1900FWA, La Plata, Argentina}
\affil{$^{12}$Facultad de Ciencias Astron\'omicas y Geof\'{\i}sicas, Universidad Nacional de La Plata (UNLP), Observatorio Astron\'omico, Paseo del Bosque, B1900FWA,\\ La Plata, Argentina}
\affil{$^{13}$Department of Astronomy and Yonsei University Observatory, Yonsei University, Seoul 03722, Republic of Korea}
\affil{$^{14}$Department of Physics and Astronomy, Research Centre for Astronomy, Astrophysics and Astrophotonics, Macquarie University, Sydney NSW 2109, Australia}
\affil{$^{15}$ Centre for Extragalactic Astronomy, University of Durham, Stockton Road, Durham DH1 3LE, United Kingdom}
\affil{$^{16}$ Institut de Ciencies del Cosmos (ICCUB), Universitat de Barcelona (IEEC-UB), E02028 Barcelona, Spain}
\affil{$^{17}$ School of Physics and Technology, Wuhan University, Wuhan 430072, China}
\affil{$^{18}$ N\'ucleo de Astronom\'ia de la Facultad de Ingenier\'ia y Ciencias, Universidad Diego Portales, Av. Ej\'ercito Libertador 441, Santiago, Chile}
\affil{$^{19}$ Institute of Astrophysics, Pontificia Universidad Cat\'olica de Chile, Av.~Vicu\~na Mackenna 4860, 7820436 Macul, Santiago, Chile}
\affil{$^{20}$ Centre for Astrophysics Research, Department of Physics, Astronomy and Mathematics, University of Hertfordshire, Hatfield, AL10 9AB, UK}
\affil{$^{21}$ Kavli Institute for Astronomy and Astrophysics, Peking University, 5 Yiheyuan Road, Beijing 100871, China}
\affil{$^{22}$ School of Mathematics and Physics, University of Queensland, Brisbane, QLD 4072, Australia}
\affil{$^{23}$ Institute of Cosmology and Gravitation, University of Portsmouth, Burnaby Road, Portsmouth PO1 3FX, UK}
\affil{$^{24}$ Department of Physics and Astronomy, University of Southampton, Highfield, SO17 1BJ, UK}
}%

\jid{PASA}
\doi{10.1017/pas.\the\year.xxx}
\jyear{\the\year}

\usepackage{aas_macros}
\usepackage{hyperref} 
\hypersetup{colorlinks,citecolor=blue,linkcolor=blue,urlcolor=blue}

\begin{document}

\begin{frontmatter}
\maketitle
\end{frontmatter}

\begin{frontmatter}
\begin{abstract}
We present an overview of the \underline{M}iddle \underline{A}ges \underline{G}alaxy \underline{P}roperties with \underline{I}ntegral Field Spectroscopy (MAGPI) survey, a Large Program on the European Southern Observatory Very Large Telescope. MAGPI is designed to study the physical drivers of galaxy transformation at a lookback time of 3--4 Gyr, during which the dynamical, morphological, and chemical properties of galaxies are predicted to evolve significantly. The survey uses new medium-deep adaptive optics aided Multi Unit Spectroscopic Explorer (MUSE) observations of fields selected from the Galaxy And Mass Assembly (GAMA) survey, providing a wealth of publicly available ancillary multi-wavelength data. With these data, MAGPI will map the kinematic and chemical properties of stars and ionised gas for a sample of 60 massive ($> 7 \times 10^{10} M_\odot$) central galaxies at $0.25 < z <0.35$ in a representative range of environments (isolated, groups and clusters). The spatial resolution delivered by MUSE with Ground Layer Adaptive Optics (GLAO, $0.6-0.8$ arcsec FWHM) will facilitate a direct comparison with Integral Field Spectroscopy surveys of the nearby Universe, such as SAMI and MaNGA, and at higher redshifts using adaptive optics, e.g. SINS. In addition to the primary (central) galaxy sample, MAGPI will deliver resolved and unresolved spectra for as many as 150 satellite galaxies at $0.25 < z <0.35$, as well as hundreds of emission-line sources at $z < 6$. This paper outlines the science goals, survey design, and observing strategy of MAGPI. We also present a first look at the MAGPI data, and the theoretical framework to which MAGPI data will be compared using the current generation of cosmological hydrodynamical simulations including {\sc EAGLE}, {\sc Magneticum}, {\sc HORIZON-AGN}, and {\sc Illustris-TNG}. Our results show that cosmological hydrodynamical simulations make discrepant predictions in the spatially resolved properties of galaxies at $z\approx 0.3$. MAGPI observations will place new constraints and allow for tangible improvements in galaxy formation theory.
\end{abstract}

\begin{keywords}
Surveys -- galaxies: evolution -- galaxies: kinematics and dynamics -- galaxies: star formation -- galaxies: stellar content -- galaxies: structure
\end{keywords}
\end{frontmatter}

\section{Introduction}
\label{sec:intro}

The question of ``nature vs. nurture'' in determining the evolution of galaxies over cosmic time is an outstanding issue in astrophysics. Nature refers to processes that are inherent to a galaxy; for example internal processes such as radial migration, gravitational instabilities, as well as energetic feedback from massive stars and super-massive black holes. Nurture instead refers to the importance of environment in shaping galaxy properties, typically through interactions with other galaxies or their host halo.  Disentangling the influence of these competing internal (nature) and external (nurture) mechanisms has proven extremely difficult, requiring detailed measurements of galaxies' internal properties (e.g. stellar and gas kinematics, chemical abundances, and star-formation histories) across a broad range of environments and lookback times \citep[see e.g.][for a review of current theoretical challenges in galaxy formation]{Naab17}.

In the nearby Universe, galaxy properties are known to correlate strongly with the properties of their host environments. The most obvious example of this correlation is in terms of galaxy morphology, where visually classified early-type galaxies are preferentially found in high-density regions \citep[i.e. the morphology--density relation,][]{Dressler80, Deeley17}. It has also been shown that galaxies in dense environments are redder (i.e. older and/or more metal-rich), more concentrated, more massive, have depleted star formation rates and lower angular momentum on average than galaxies in the field \citep[e.g.][]{Kauffmann04,Blanton05,Cooper06,Skibba09,Davies19}. Residual stellar populations trends with environment were shown to persist even when accounting for stellar mass \citep{Liu16,Scott17}.

The extent to which these correlations represent systematic differences in \emph{intrinsic} galaxy properties, or are instead a reflection of the processes acting \emph{within} high-density environments, remains unclear.  \citet{Peng10} argued that stellar mass is the primary driver of galaxy colour in massive galaxies regardless of their host environment at $z\approx0$, with environmental processes only becoming relevant at lower stellar masses. 
%\citet{Brough17} used two-dimensional kinematic data from the SAMI galaxy survey \citep{Croom12} to show that there is similarly very little dependence of kinematic morphology (i.e. fast vs. slow rotators) on environment at fixed stellar mass \citep[also see][]{Greene17,Veale17}. 
While ``semi-analytic'' models suggest that there should be a correlation between the formation histories of galaxies and their host dark matter haloes \citep[e.g.][]{Kauffmann95,DeLucia06}, such signatures remain confused in observational data \citep[see, e.g.][]{Thomas05,Thomas10,Cooper10, Brough13, Davies19}. Nevertheless, there is clear evidence that numerous physical processes can and do affect galaxy evolution inside group and cluster environments (see the review of \citealp{Boselli06}), including interactions between galaxies and the intra-cluster medium (e.g. ram-pressure and viscous stripping, e.g. \citealt{vanderWel10}), galaxy--galaxy mergers \citealt{Oh18,Oh19}, and flybys (so-called ``harassment'', e.g. \citealt{Robotham14}; \citealt{Davies15}).

Ultimately, the variety of timescales over which internal vs. external processes are expected to act complicates the interpretation of observations at a single (recent) epoch, and motivates the incorporation of higher redshift data to break the degeneracy between different evolutionary pathways. Initial investigations of galaxy morphology at $z \gtrsim 1$ using optical \textit{Hubble Space Telescope} (HST) imaging revealed an abundance of clumpy and irregular morphologies typically associated with gas-rich mergers \citep[e.g.][]{Driver95a,Driver95b,Glazebrook95,Baugh96}.  However, subsequent multi-wavelength observations have demonstrated that the overall picture of galaxy evolution since $z \sim 1-3$ is complex.  Despite their disturbed appearance at optical wavelengths (rest-frame ultraviolet), studies based on deep near-infrared imaging have shown that normal star-forming galaxies at nearly every epoch have light profiles that are well described by an exponential disk \citep{Wuyts11}. This apparent regularity in structure is supported by resolved studies of ionised gas kinematics at $z \gtrsim 1$, which show that the majority of galaxies are consistent with marginally stable disks and short dynamical times \citep{Wisnioski15,Stott16,ForsterSchreiber18, Ubler19}, albeit significantly truncated in size when compared to local discs \citep{Trujillo05, vanderWel14}.

Extending lookback studies to include stellar properties---in particular resolved kinematics---is more difficult on account of the stellar body being significantly fainter. Nevertheless, significant progress has been made through a combination of deep long-slit observations and targeted follow-up of lensed high-redshift sources, which suggest that the rotational support prevalent among star-forming galaxies at $2 < z < 3$ persists even as their star formation is ultimately quenched \citep[e.g.][]{Toft17,Newman18}.  Even at $z\approx0.8$, the degree of rotational support observed in massive quiescent galaxies is a factor of $\sim$2 higher than at $z=0$ \citep[e.g.][]{Bezanson18}.

That significant kinematic evolution is inferred at $z < 1$ should not be surprising: even though the merger rate decreases significantly with decreasing redshift \citep[e.g.][]{Conselice14,Robotham14,LopezSanjuan15,Mundy17}, the reduced rate of cosmological accretion and corresponding reduction in gas available for star formation mean that galaxies have less chance to ``recover'' angular momentum following a merger event \citep{Penoyre17,Lagos18b}. Repeated gas-poor interactions therefore provide an efficient (albeit not exclusive) mechanism to drive kinematic and morphological transformation of the galaxy population, however understanding when and where such transformations take place requires tracking the detailed kinematic properties of both gas and stars over significant stretches of cosmic time.

Local Integral Field Spectroscopy (IFS) studies to date have made extensive use of the stellar spin parameter in order to kinematically classify galaxies. This spin parameter is an observational proxy of the intrinsic spin of galaxies first suggested by \citet{Emsellem07}, and defined as:
\begin{equation} 
\lambda_{\rm r} \equiv \langle R|V|\rangle/\langle R\sqrt{V^2+\sigma^2}\rangle,
\end{equation}
where $V$, $\sigma$, and $R$ are the normalised recession velocity, velocity dispersion and circularised galactocentric radius at a given projected position. As a simple probe of the overall dynamical state of a galaxy, $\lambda_{\rm r}$ is a popular diagnostic parameter that is readily derived from spatially resolved spectroscopy.

\begin{figure}
\includegraphics[clip,width=0.48\textwidth]{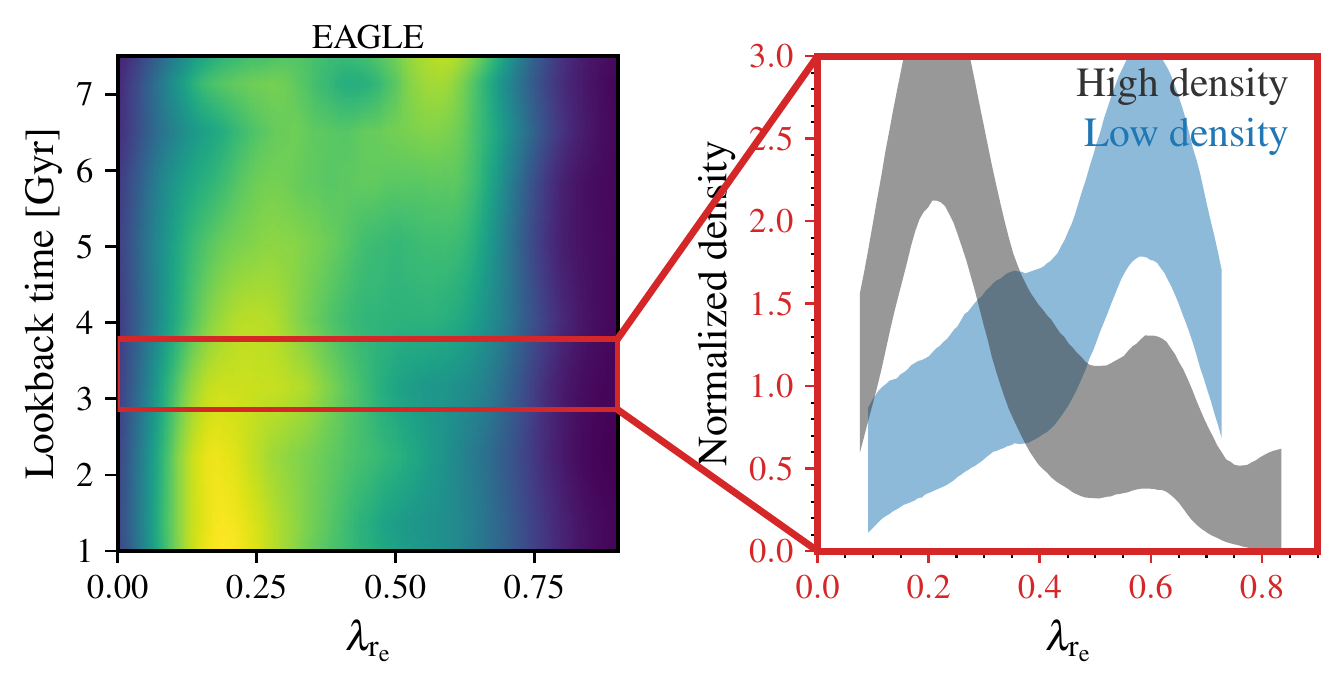}
\includegraphics[clip,width=0.48\textwidth]{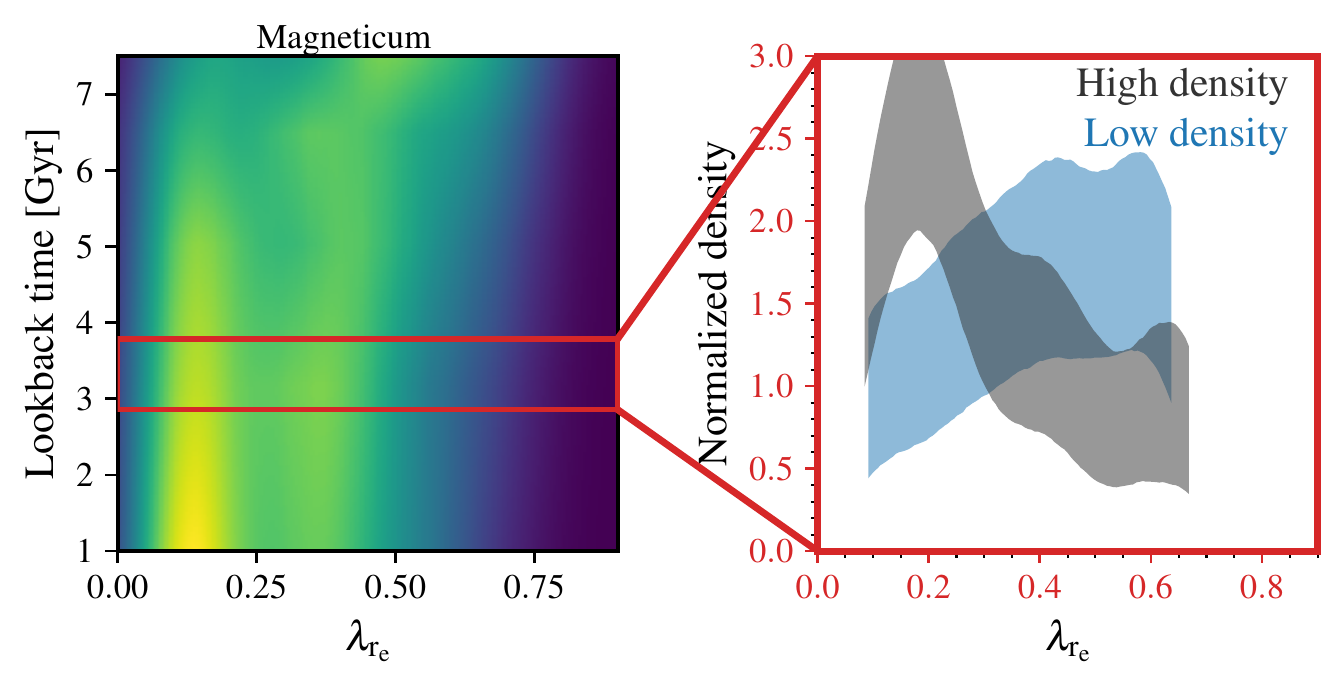}
\includegraphics[clip,width=0.48\textwidth]{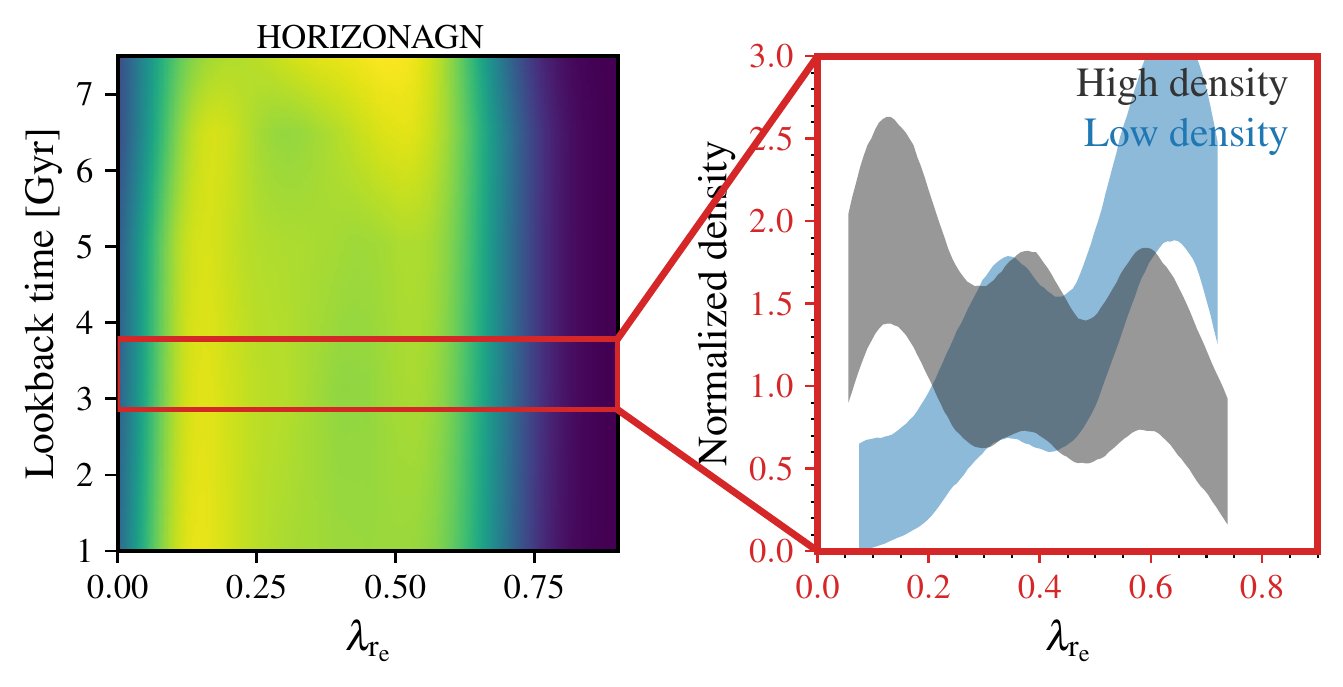}
\caption{{\it Left panels:} Distribution of galaxies in the $\lambda_{\rm r_e}$-lookback time plane for the {\sc EAGLE} (top panel), {\sc Magneticum} (middle panel) and {\sc HORIZON-AGN} (bottom panel) hydrodynamical simulations. We select MAGPI-like primary targets in the three simulations (which we simply select as those with stellar masses $>10^{10.8}\,\rm M_{\odot}$), and randomly sample those to match the number of expected MAGPI primary targets (see $\S$~\ref{section:resultsims} for more details on the sampling). The colour shows the linear number density, with yellow indicating higher concentration of galaxies. {\it Right panels:} probability density function of $\lambda_{\rm r_e}$ in high and low density environments, defined as the top and bottom thirds of the host halo masses of galaxies, respectively (the exact value in halo mass of these thresholds therefore depends on the simulation; see for $\S$~\ref{subsec_simsresultskinematics} for details). The uncertainty regions are computed based on the expected number of MAGPI galaxies. All simulations predict significant transformation in $\lambda_{\rm r_e}$ of massive galaxies at $z<1$. At the redshift range of MAGPI (red box in the left panels) the simulations predict different levels of environmental effects, which will be tested by our survey. See $\S$~\ref{subsec_simsresultskinematics} for a more in-depth discussion of this figure.}\label{fig:lambdaRevol}
\end{figure}

One key finding of local IFS studies is that galaxies can be divided into two main dynamical families according to their position in $\lambda_{\rm r_e}-\epsilon$ space, where $\lambda_{\rm r_e}$ is $\lambda_{\rm r}$ measured at the effective (half-light) radius, $r_{\rm e}$, and $\epsilon$ is the projected ellipticity. Two dynamical classes separate in spin for a given projected ellipticity: fast-rotators (high $\lambda_{\rm r_e}$) and slow-rotators (low $\lambda_{\rm r_e}$). The division between these two common classes continues to be nuanced \citep{Emsellem07,Emsellem11,Cappellari16, Graham18, vandeSande20}. The origin of this possible bimodality is still unclear, with theoretical simulations and detailed observational studies finding multiple possible formation pathways for the rarer slow-rotator population (e.g.  \citealt{Khochfar11,Penoyre17,Lagos18b,Schulze18,Krajnovic20,WaloMartin20}, also see Fig.~\ref{fig:lambdaRevol}). 
%NEEDS SPECIFIC REFERENCES FIRST: Indeed, successive IFS studies have revealed and confirmed that the diversity of galaxy properties seen today are consistent with a complex interplay of accretion, gas-poor and gas-rich mergers, and a variety of star formation histories. 

\begin{figure*}
\centering
\includegraphics[width=17.5cm]{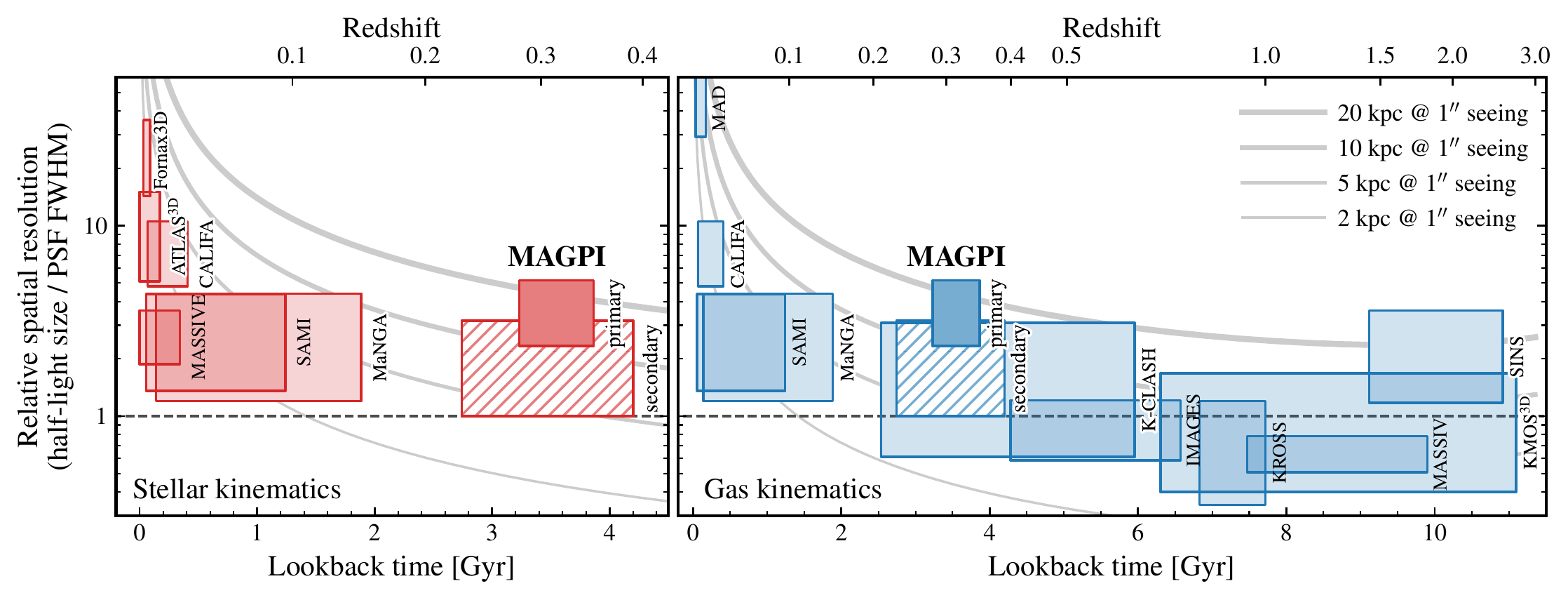}
\caption{Comparison of the MAGPI spatial resolution with that of other dedicated IFS surveys focused on stellar (left panel) and gas (right panel) kinematics. Shaded regions indicate the typical space occupied by surveys in terms of lookback time and spatial resolution, defined here as the ratio of galaxy half-light size relative to the PSF FWHM. We compare data from the MAGPI primary and secondary samples (see \S~\ref{section:sample}) to that of other IFS surveys, including SAMI \citep{Croom12}, MaNGA \citep{Bundy15}, MASSIVE \citep{Ma14}, CALIFA \citep{Sanchez12}, Fornax3D \citep{Sarzi18}, ATLAS$^{\rm 3D}$ \citep{Cappellari11a}, MAD \citep{ErrozFerrer19}, K-CLASH \citep{Tiley20}, IMAGES \citep{Yang08}, MASSIV \citep{Contini12}, KMOS$^\mathrm{3D}$ \citep{Wisnioski15,Wisnioski19}, KROSS \citep{Stott16}, and SINS/zC-SINF \citet{ForsterSchreiber18}. Background curves show how galaxies with a fixed physical sizes (as indicated), appear in this parameter space for observations taken in 1 arcsec FWHM seeing conditions.}\label{fig:spatialres}
\end{figure*}

%The importance of gas + stars.
 To dissect the evolutionary pathways that transformed the primarily disky/irregular systems at high redshift into today's rich morphological mix of galaxies, it is essential to measure both the stars and ionised gas simultaneously in a range of environments.
%A dedicated observational campaign that can spatially map stellar \emph{and} ionised gas properties of galaxies beyond 2 Gyr of cosmic time and across a range of environments is required to disentangle the role of various physical processes in shaping galaxies. 
Because such IFS observations are time intensive, available data so far have been limited to small samples or lower-resolution slit spectra along specific position angles \citep[e.g.][]{Moran07, vanderMarel07, vanderWelvanderMarel08, vanderWel16} $-$ providing limited constraints for detailed theoretical models of galaxy evolution.  \citealt{Guerou17} simultaneously study IFS stellar and ionised gas kinematics in a limited sample of 17 galaxies beyond the redshifts already probed by local studies (i.e. $z>0.15$).
IFS is the only technology that allows for stellar and gas phase properties to be fully and simultaneously mapped. The absence of a substantial \textit{stellar} IFS dataset beyond $z\sim0.15$; and until recently, ionised gas IFS data between $0.15<z<0.70$ \citep[][see Fig. \ref{fig:spatialres}]{Carton18,Tiley20,Vaughan20}; greatly limited our understanding of galaxy evolution during the Universe's middle ages when morphology, angular momentum and star formation activity evolve rapidly, with environment playing a key role \citep[see Fig. \ref{fig:lambdaRevol} and e.g.][]{Peng10, Papovich18,Choi18}.

%\subsection*{The MAGPI Survey}

%\newline

%MAGPI is an open collaboration. Membership is achieved through submitting an unique project idea. Individual projects are protected over an appropriate time frame with team members with overlapping interests encouraged to collaborate.

%All members of the MAGPI team have agreed to follow the MAGPI Survey Policy and the Astronomical Society of Australia's Code of Conduct\footnote{\url{https://asa.astronomy.org.au/membership/conduct-and-ethics/}}. The latter ensures that all members know they are expected to behave professionally. The Survey Policy details the survey management model, procedure for engaging new members, authorship guidelines and ensures that all contributions are fairly rewarded.

%This decentralised approach allows all members to make meaningful contributions to the survey according to their individual expertise and availability.

%Given the broad range of geographical locations of team members, and the requirement for effective communication associated with distributed leadership, the team uses a range of tools to keep members engaged and to encourage ongoing discussion. Important decisions and notes are broadly shared in all important meetings to ensure members who are unable to attend are kept up-to-date. This has the additional benefit of ensuring transparency and accountability.

This paper presents the  \underline{M}iddle \underline{A}ges \underline{G}alaxy \underline{P}roperties with \underline{I}FS (MAGPI) survey. It is divided as follows: in \S~\ref{section:sciencegoals}, we describe the Survey and science goals. The sample description, survey design, observing strategy and data handling can be found in \S~\ref{sec:data}. \S~\ref{section:results} showcases early observational and theoretical results, while a brief summary can be found in \S~\ref{section:summary}.

For observational results and unless otherwise stated, we assume a $\Lambda$CDM cosmology with  $\Omega_{\rm m}=0.3$, $\Omega_{\lambda}$~$=$~$0.7$ and $H_0=70$ km s$^{-1}$ Mpc$^{-1}$. We use AB magnitudes throughout \citep{Oke1983}, and stellar masses have been derived assuming a \citet{Chabrier03} stellar initial mass function.

\section{The MAGPI Survey and Science Goals}\label{section:sciencegoals}

Until now, there has not been a dedicated observational campaign that can spatially map stellar \emph{and} ionised gas properties of galaxies beyond 2 Gyr lookback time, as is necessary to disentangle the role of various physical processes in shaping galaxies (see Fig. \ref{fig:spatialres}).
Aiming to close the important gap in IFS gas studies and double the evolutionary window of local IFS studies of stars, we present the MAGPI Survey, a VLT/MUSE Large Program (Program ID: 1104.B-0536)
that is currently gathering observations of resolved gas and stars at $z=0.25-0.35$ in 60 ``primary target'' galaxies ($M_\ast > 7\times10^{10} M_\odot$) and their $\sim$100 satellites in a range of environments, including isolated galaxies. The sample is achieved through dedicated $56\times4$ hours on-source observations with ground layer adaptive optics (GLAO) on VLT/MUSE ($1\times1$ arcmin field-of-view), in combination with 2 legacy archive fields Abell 370 and Abell 2477; see Table \ref{tab:targetlist2} and Fig. \ref{fig:fieldsgallery}). The survey is designed to reveal the physical processes responsible for the rapid transformation of galaxies at the relatively unexplored intermediate redshift regime.

%Survey management strategy.
MAGPI is led through a distributed leadership model \citep{Pilkiene18} with a leadership team currently composed of 4 equal Principal Investigators (PIs): Foster, Lagos, Mendel, and Wisnioski (in alphabetical order). All PIs contribute to the management and leadership of the survey. Major decisions are made by consensus through discussion. 
Team members are encouraged to contribute to the survey management and effort through four working groups: the Master Catalogue, Emission Lines, Absorption Lines and Theory Working Groups. 
General information about the survey, including how to contact or join the MAGPI team, can be found on the survey website: \url{https://magpisurvey.org}.

MAGPI will map the detailed properties of the stars \emph{and} ionised gas for galaxies in a range of halo masses ($M_{\rm halo}$) with lookback time of 3--4 Gyrs. The main goal of MAGPI is to reveal and understand the physical processes responsible for the rapid transformation of galaxies at intermediate redshifts by:
\begin{itemize}
\item detecting the impact of environment (\S~\ref{sec:goal1});
\item understanding the role of gas accretion and merging (\S~\ref{sec:goal2});
\item determining energy sources and feedback activity (\S~\ref{sec:goal3});
\item tracing the metal mixing history of galaxies (\S~\ref{sec:goal4}); and
\item producing a comparison-ready theoretical dataset (\S~\ref{sec:theorycase}).
\end{itemize}
In addition to the main science cases, MAGPI will enable serendipitous  higher redshift emission-line (e.g., [OII] emitters at $0.35 < z < 1.50$, \citealt{Herenz17}) and Lyman-$\alpha$ emitter ($2.9 <z< 6.0$, \citealt{Herenz19}) science.

%%%-----STARS CASE (ANGULAR MOMENTUM) -------
\subsection{Detecting the impact of environment}\label{sec:goal1}
%%%------------------------

%The most massive galaxies in the Universe tend to be in the centre of the deepest potential wells such as the centres of galaxy clusters. The fraction of early to late-type galaxies increases with stellar mass \citep[e.g.][]{Kauffmann04,Vulcani11,Kawinwanichakij17}, and hence environmental density, a trend known as the morphology-density relation \citep[e.g.][]{Dressler80, Goto03, Deeley17}. Massive galaxies also tend to have a higher proportion of slow rotators, which naturally leads to the kinematic-morphology relation \citep[e.g.][]{Cappellari11b,Scott14,Fogarty15}. However, it is unclear what the respective roles of environment and secular processes are in transforming galaxies with recent studies suggesting that differences in the kinematic morphology of galaxies can be ascribed primarily to stellar mass trends \citep[e.g.][]{Greene17,Brough17}. On the other hand, there is clear evidence that environment plays an important role in quenching star formation \citep[e.g.][]{Skibba09, Bamford09, Peng10, Smith12, Cappellari16}.

To resolve the role of external processes (i.e. nurture) in transforming galaxies, MAGPI will explore the effect of local vs large-scale environmental density at a key epoch. Simulations \citep[e.g.][]{Penoyre17, Lagos18b} suggest that large-scale environmental trends should be more pronounced at intermediate redshifts, where environment is predicted to play a more active role in galaxy formation. Fig.~\ref{fig:lambdaRevol} shows the $\lambda_r$ distributions as a function of cosmic time for a randomly selected sample of $60$ massive galaxies (stellar masses $\ge 10^{10.8}\,\rm M_{\odot}$) at each epoch (left) and split into environment bins (right) of $20$ massive galaxies each for three different cosmological simulations, \textsc{EAGLE}, \textsc{Magneticum} and \textsc{HorizonAGN}, each showing very different evolutionary patterns at these redshifts (see $\S$~\ref{section:resultsims} for details).

The spatial resolution, data quality and availability of panchromatic ancillary data, allow for a detailed, quantitative comparison between MAGPI and both local observations and simulations. By targeting galaxies at the critical epoch during which the impact of evolutionary processes on galaxy dynamics are likely maximised, MAGPI data give us the best opportunity to identify external formation pathways for massive central galaxies and their satellites in different environments.

\subsection{Understanding the role of gas accretion and merging.} \label{sec:goal2}

%A bit of intro/motivation
Repeated dynamical interactions can qualitatively reproduce the observed differences in morphology and $\lambda_{\rm r}$ required to turn present-day spirals into early-type galaxies \citep{BekkiCouch11}. Accretion of gas from either gas-rich mergers or external accretion can lead to the (re-)formation of a disc, destruction of spiral arms, and overall spin-up of the system \citep[e.g.][]{Dubois16,Sparre17,Lagos18a}. The frequency and impact of both processes are known to evolve over cosmic time \citep{Rodriguez-Gomez15,Wright20}. Some theoretical studies suggest that gas poor mergers are one of the main drivers in producing the slowly rotating galaxies we observe today (\citealt{Naab14}; \citealt{Schulze18}, \citealt{Lagos18a}, but see e.g. \citealt{Kobayashi04}; \citealt{Cox06}; \citealt{Taranu13}; \citealt{Penoyre17}), and because their frequency is expected to increase at $z<1$ \citep{Lagos18a}, we expect the last few billion years to be critical in building the diversity observed in galaxies in the local Universe.

The epoch of $0\le z\le 1$ is also known as the ``disc settling'' epoch where galaxies that continue to accrete gas and form stars can efficiently build up their specific angular momentum \citep{Kassin12,Simons17,Lagos17,Ma17,Wisnioski19}. This is a natural result from hierarchical cosmologies, in which the specific angular momentum of the accreted gas is expected to increase with time \citep{Catelan96,Teklu15, El-Badry18}. The latter implies that the later the accretion and star formation, the more likely the galaxy will have a high spin at the present day. Quantifying the interplay between mergers and gas accretion, when both processes are thought to be significant, is critical to understanding morphological and chemical transformations. 

With MAGPI and existing low-redshift IFS surveys, we will establish the evolution of the role of mergers and gas accretion in transforming galaxies across halo mass and the evolution of such processes over the last 4 Gyrs. %For example, asymmetries in kinematic maps, measured through kinemetry \citep{Krajnovic06} can reveal signatures of past major mergers seen in stellar kinematic maps \citep[e.g.][]{Oh16} or recent accretion events seen in asymmetries of gas kinematic maps \citep[e.g.][]{BarreraBallesteros15,Bloom18}. We will measure the relative alignment of the stellar and gas components to identify clear misalignments \citep[e.g.][]{Davis11,Bryant19}, a tell-tale signature of recent gas accretion \citep{vandevoort15}, as a function of environment for all galaxies where both gas and stars are resolved ($\sim 100$ galaxies expected).

\begin{figure*}
\centering
\includegraphics[width=\textwidth, trim=0mm 12mm 0mm 0mm]{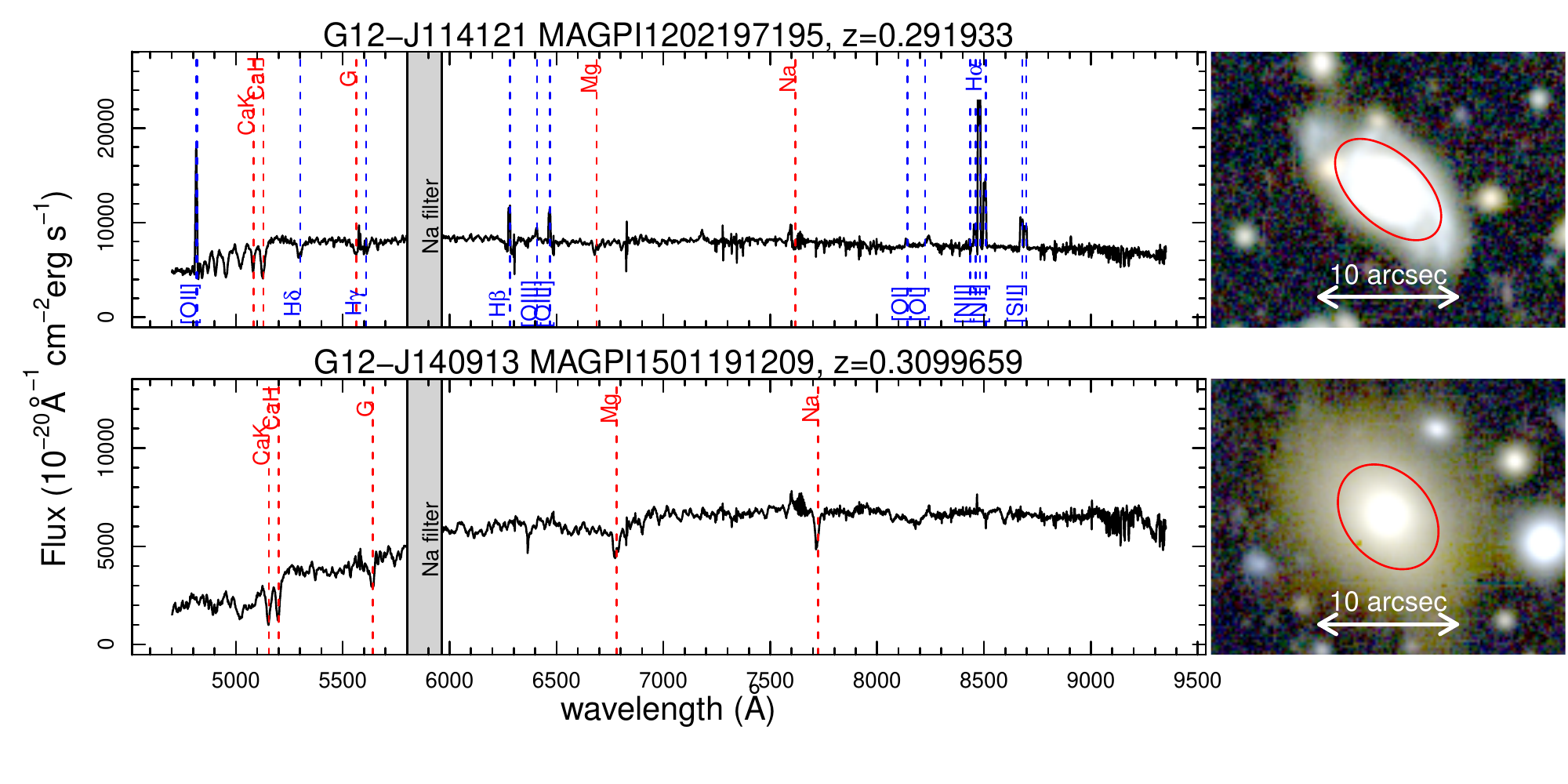}
\caption{Left: Example observed $2r_e$ aperture spectra for the central galaxies of fields G12-J114121 (MAGPI ID: 1202197195, top) and G15-J140913 (MAGPI ID: 1501191209, bottom). Both show common absorption (red) lines, while the former also shows emission (blue) lines. The grey shaded area shows the wavelength range blocked by the sodium laser filter. Right: Synthetic $g_{\rm mod}ri$-colour images of the respective galaxies showing the $2r_e$ aperture radius.}\label{fig:spectra}
\end{figure*}

\subsection{Determining energy sources and feedback activity} \label{sec:goal3}

% older texts (feel free to restore)
%The transformation from star-forming to passive galaxies in different environments leaves distinct imprints on the inter-stellar medium. Resolved observations of the inter-stellar medium are essential to determine the energy sources driving star formation and feedback, thus regulating the dynamical assembly of galaxies \citep[e.g.][]{Ho15, Belfiore17}. With AO-aided observations, MAGPI delivers high S/N emission-line maps of star-forming galaxies extended to $\sim2-3r_e$ with $\sim2.7$ kpc resolution. Properties (color, mass, size) of the star-forming galaxies inthe the sample are shown in Fig. \ref{fig:sampleSelect}. The wealth of emission lines ([OII]$\lambda\lambda3727$ to [SII]$\lambda\lambda6717,31$), available only below $z=0.38$, will simultaneously trace active galactic nuclei (AGN) as well as resolved star formation and ISM properties such as metallicity, shock, ionisation parameters, and electron density \citep{Yuan12, Davies14, Ho15}. With these tools MAGPI enables the investigation of the role of feedback-driven processes (e.g. AGN; inside-out quenching) and external processes (e.g. ram pressure stripping; outside-in quenching) in producing the present-day passive galaxy population \cite{Guo19}.

Stars and active galactic nuclei (AGN) are the main energy sources that produce the spectral energy distribution (SED) and emission lines of galaxies \citep[see][for a recent review]{Kewley19}.  The radiation and kinetic energy from stars and AGN are consumed and re-processed in and through the interstellar medium (ISM) via a rich set of physical processes. Feedback is key amongst these processes, including photoionisation, collisions, shocks, winds, and outflows; all of which can significantly impact the star formation history of galaxies.
Feedback processes are considered critical in quenching star formation in massive galaxies and accounting for the observed stellar mass function \citep[e.g.,][]{Man18}. However, a concrete picture of how feedback by energetic sources modulates the evolution and growth of massive galaxies remains elusive in both theory and observation \citep{Fabian12,Naab17}. %Resolved observations of the ionised gas are essential to determine the energy sources driving star formation and feedback, thus regulating the dynamical assembly of galaxies \citep[e.g.][]{Ho15, Belfiore17}.  

The key to clearly delineate energy and feedback sources in galaxies is to spatially diagnose and distinguish them.
%Rest-frame  optical emission lines are extremely useful diagnostics of the energy sources \citep{Baldwin81,VO87,Kewley06}.%With AO-aided observations, MAGPI delivers high signal-to-noise (S/N) rest-frame optical emission lines of star-forming galaxies extended to $\sim2-3r_{\rm e}$ with $\sim1.8-2.7$ kpc resolution. %Properties (color, mass, size) of the star-forming galaxies in the sample are shown in Fig. \ref{fig:sampleSelect}. 
%The large optical range of emission lines ([OII]$\lambda\lambda3727$ to [SII]$\lambda\lambda6717,31$), available for our primary targets ($0.27 \le z \le 0.39$, described in \S~\ref{section:sample}, example spectra in Fig.~\ref{fig:spectra}), ensures accurate diagnostics of regions that trace AGN and star formation activities. 
With MAGPI, we will simultaneously decode the feedback signatures from the resolved star formation rate, dust attenuation, and ISM properties such as metallicity, shock velocity, ionisation parameters, and electron density \citep{Yuan12, Davies14, Ho15} using rest-frame optical emission line diagnostics \citep{Baldwin81,VO87,Kewley06, Poetrodjojo21}. With MAGPI, environmental and in-situ quenching mechanisms will be correlated with the spatial distribution of star formation at redshift $z\approx0.3$ \citep[also see][]{Vaughan20} and compared to local trends \citep[e.g.][]{Schaefer19,Bluck20} to identify evolution in the prominence of various quenching mechanisms.

%The spatial resolution, spectral coverage, and representative range of environments covered by the MAGPI fields, constitute an unprecedented sample at $z\approx0.3$ to study the relative importance of various stellar and AGN feedback processes in transforming galaxies. 

\subsection{Tracing the metal mixing history of galaxies} \label{sec:goal4}

%In addition to morphological transformations, accretion and merging are expected to have a significant impact on the enrichment history of and metal distribution within galaxies. 
Radial metallicity gradients of both gas and stars provide temporal snapshots of a galaxy's chemical history. Recent chemodynamical cosmological simulations show that a joint picture of stellar and gas metallicity gradients provide one of the most stringent constraints on the mass assembly history of both late- and early-type galaxies \citep{Taylor17, Tissera18}. Across cosmic time, the predictions for both stellar and gas metallicities, show sensitive dependence on the history of merger events, AGN feedback and star formation. 

This dependence is reflected in the large scatter seen in local gas metallicity gradient observations \citep{Belfiore17, SanchezMenguiano16} and beyond $z\sim0.2$ \citep{Queyrel12,Stott14, Wuyts16, Carton18, ForsterSchreiber18}. 
Notably, the largest scatter in slopes is predicted beyond $>1r_{\rm e}$ in massive galaxies $2-6$ Gyrs ago; reflecting that a broad range of accretion histories, kinematics, and feedback mechanisms are at play \citep{Ma17}.  \citet{Collacchioni20} showed that even within $1r_{\rm e}$, gas accretion clearly affects the slope of gas metallicity profiles in {\sc EAGLE} simulations. These simulations also predict that AGN play an important role in setting radial metallicity gradients, with resolved mass vs gas-phase metallicity relations turning over under the influence of AGN feedback \citep{Trayford19b}.

With simultaneous gas \textit{and} stellar metallicity measurements at $z=0.3$, these models can now be confronted with joint observations at higher redshifts for the first time. In other words, MAGPI will establish the first comprehensive dataset at intermediate redshift to test chemodynamical models using stellar and gas metallicity gradients, along with a detailed study of how gas and stellar metallicity gradients vary with galaxy and environment properties.

\subsection{Producing a comparison-ready theoretical dataset}\label{sec:theorycase}

%An important strategy we are adopting in MAGPI is the close connection with simulations, which serves two main purposes. 
The MAGPI survey has close connections with a variety of cosmological simulations. This is an important element for two main reasons.
Firstly, simulations provide the necessary context for our sample selection and the analysis of our observational results.
Simulations equip the team with a resource to quantify the completeness of the environment sampling and spectroscopic completeness.
%Particularly, we are interested in understanding how well we are sampling environment given the spectroscopic completeness we obtain, and whether a population of galaxies is expected to be missing due to flux limitations, among other. 

Secondly, MAGPI observations allow us to test the wealth of predictions from large-scale galaxy simulations as well as from analytic and semi-analytic models. For this, it is essential to explore a suite of simulations to provide us with predictions that appear robust to the details of galaxy formation modelling and predictions that are highly dependent on those details. The main aims are to pin-point areas that require revision in simulations, and to understand whether or not the modelling of specific physical processes (e.g. stellar or AGN feedback) implemented in some simulations better capture the observations compared to other plausible models of the same physical process. The latter is key to move from a qualitative understanding of galaxy formation to a quantitative one.

In this and future work, we make use of existing cosmological hydrodynamical simulations and retrieve data from {\sc EAGLE} \citep{Schaye15,Crain15}, {\sc Magneticum} \citep{Teklu15,Schulze18}, {\sc HORIZON-AGN} \citep{Dubois16}, {\sc Illustris-TNG100} \citep{Pillepich18, Naiman18, Springel18, Nelson19b} and 
the chemo-dynamical simulation of \citet{Taylor15} and \citet{Taylor17}, henceforth TK15.
As more simulations become available we will continue to increase our library of predictions. An important aspect of our strategy is to have experts on all these simulations as part of our team, in order to have first-hand knowledge of the technical details of each of them. In \S~\ref{section:datasims}, we provide a brief description of the simulations that are currently part of our suite, while \S~\ref{section:resultsims} showcases early theoretical results.

\section{Data}\label{sec:data}
The MAGPI sample (\S~\ref{section:sample}), observing strategy (\S~\ref{section:observations}), data processing (\S~\ref{section:reduction}) and theoretical dataset (\S~\ref{section:datasims}) are designed and implemented to optimally address the survey goals described in \S~\ref{section:sciencegoals}.

\subsection{Sample selection and survey design} 
\label{section:sample}

\begin{figure*}
\centering
\includegraphics[scale=0.9]{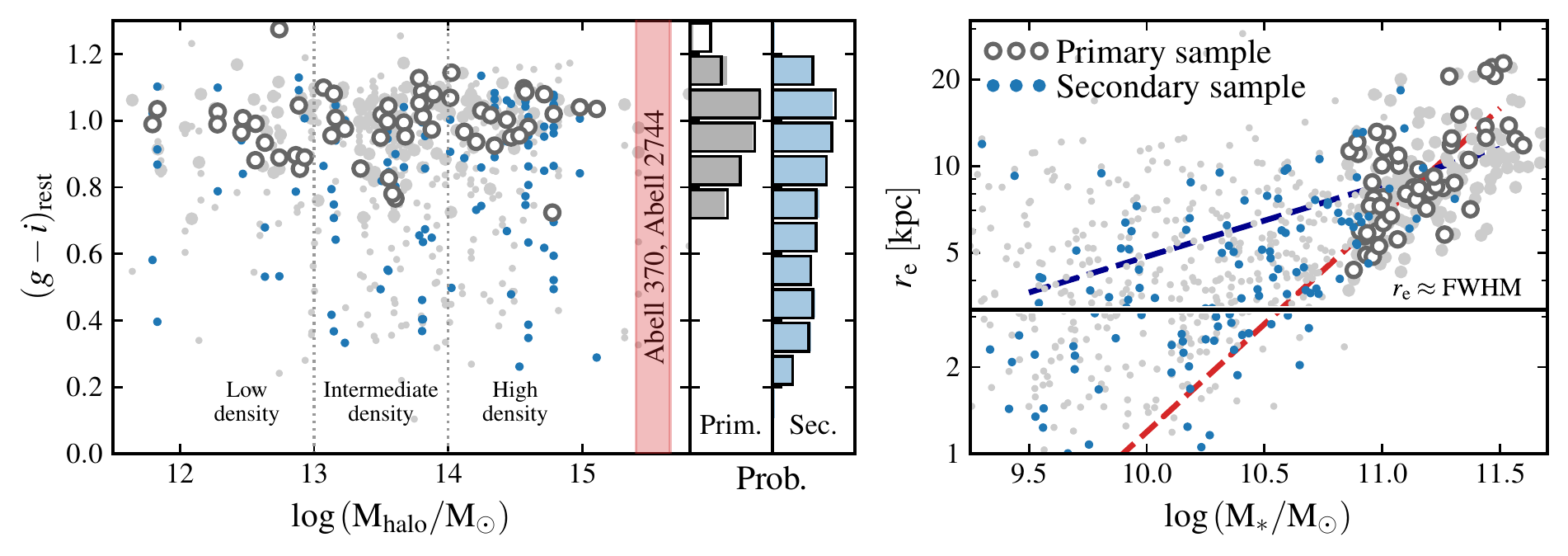}
\caption{Illustration of the final MAGPI target selection. \emph{Left panel}: the distribution of MAGPI targets in terms of $g-i$ colour and dark matter halo mass.  Open circles indicate primary targets, while filled (blue) circles identify secondary galaxies having photometric redshifts within $\Delta z = 0.03$ of the primary target (see \S~\ref{section:sample}). Background (grey) points show the distribution of galaxies of primary and secondary galaxies in the parent sample (large and small circles, respectively). The right sub-panels show the corresponding colour histograms for the primary and secondary samples, where the parent sample is shown as filled, and the final MAGPI sample is shown as open. Primary targets were selected to sample the full observed range of both environment and colour. \emph{Right panel}: the distribution of MAGPI targets in terms of half-light size and stellar mass. Symbols are the same as in the left panel. The solid horizontal line indicates where galaxies are nominally resolved, (i.e. FWHM $\approx$ $r_\mathrm{e}$). For comparison, dashed lines show the size--mass relation for star-forming (dark blue) and passive (red) galaxies as derived by \citet{vanderWel14}. While primary targets are resolved by multiple MUSE resolution elements regardless of star-formation rate, resolved information for secondary galaxies is biased towards star-forming galaxies.}\label{fig:sampleSelect}
\end{figure*}

The MAGPI science goals require that we derive spatially-resolved stellar kinematics and structural properties for galaxies spanning a range of morphology, star-formation properties, and environment. This naturally pushes us towards selecting targets from existing surveys with substantial multi-wavelength imaging and well-characterised environmental metrics. Based on bootstrap samples drawn from the EAGLE $\lambda_{\rm r_e}$ PDFs shown in Fig.~\ref{fig:lambdaRevol}, we require a minimum of 60 massive central galaxies (20 in each of the 3 environment bins) in order to detect the difference in the shape (skewness and median) in the low and high density $\lambda_{\rm r_e}$ distributions predicted by cosmological simulations at a 95 percent confidence level (99.7 percent confidence would require $\gtrsim130$ massive galaxies). We define a ``central'' galaxy as a galaxy which dominates its environment. As such, isolated galaxies are considered centrals for our purposes.

Primary MAGPI targets were drawn from the Galaxy and Mass Assembly survey \citep[GAMA;][]{Driver11,Liske15,Baldry18}.  GAMA conducted extensive spectroscopic observations covering a total of 250 deg$^2$ across five fields (G02, G09, G12, G15, and G23).  Along with 21-band photometric data spanning from the ultraviolet to the far-infrared \citep{Driver16}, the high spectroscopic completeness of GAMA targets ($\sim98$ percent at $m_r \leq 19.8$) ensures a robust characterisation of environment in terms of both near-neighbour density \citep[e.g.][]{Brough13} and dark matter halo mass \citep[e.g.][]{Robotham11}. At $z=0.3$, the limiting magnitude of $m_r = 19.8$ used to define the GAMA spectroscopic sample corresponds to a stellar mass of $\log (M_*/M_\odot) \approx 11$.

We first identified potential targets in the GAMA G12, G15, and G23 fields with spectroscopic redshifts, $z_\mathrm{spec}$, in the range $0.28 \leq z_\mathrm{spec} \leq 0.35$ and photometrically-derived stellar masses, $M_\ast$ \citep{Taylor11}, greater than $7\times10^{10}~M_\odot$. The former selects galaxies in our redshift range of interest around $z \approx 0.3$, while the latter ensures that all primary targets will be sampled by multiple MUSE resolution elements within their half-light radii. This initial pool of 209 objects was further culled based on the availability of suitably bright ($m_R \leq 17.3$) tip-tilt stars within the GALACSI technical field, which were identified by a cross-match with Gaia DR2 \citep{Gaia18}, resulting in 95 potential targets.

Selection of the final MAGPI sample was carried out based on the requirement that galaxies uniformly sample a range of environments (including isolated galaxies) and colours.  In Fig. \ref{fig:sampleSelect} we show the distribution of selected targets in terms of rest-frame $g-i$ colour and dark matter halo mass (as derived by \citealp{Robotham11}). We select a total of 56 massive galaxies from GAMA, with a remaining 4 galaxies drawn from MUSE archival observations of Abell 370 (Program ID 096.A-0710; PI: Bauer) and Abell 2744 (Program IDs: 095.A-0181 and 096.A-0496; PI: Richard) to ensure data coverage up to the highest halo masses; the final sample covers a halo mass range spanning $11.35 \leq \log(\mathrm{M_{halo}}/\mathrm{M_\odot}) \leq 15.35$. KiDS $i$-band cutouts for the 56 GAMA target fields are shown in Fig. \ref{fig:fieldsgallery}.

In addition to providing spatially-resolved spectroscopic data for the primary galaxy sample described above, the large physical extent of the MUSE field-of-view at $z \sim 0.3$ ($\sim$270 kpc) also provides dense spectroscopic sampling of the primary galaxy's host environment. The distribution of these neighbouring objects (henceforth referred to as ``secondary'' objects) in terms of colour, size and stellar mass is shown in Fig. \ref{fig:sampleSelect}. Based on GAMA photometry, we expect as many as 150 secondary galaxies for which MAGPI observations will provide spectra at S/N $> 5$ \AA$^{-1}$, with $\sim$100 of those being resolved by multiple seeing elements within their half-light radii. Secondary objects enable the robust characterisation of environment, which is central to the MAGPI science goals (\S~\ref{section:sciencegoals}).

The depth and breadth of ancillary data for MAGPI fields available mainly through the GAMA Survey enables new areas of scientific investigations. In addition to refining environmental metrics, pushing the completeness of GAMA \citep{Robotham11}, MAGPI can produce extremely deep satellite stellar mass functions for the targeted GAMA groups. 

\subsection{Observing strategy}
\label{section:observations}

Observations for MAGPI are carried out in service mode and in dark time, starting in ESO Period 104, and being a large program, will continue until completion. MUSE is used in the wide-field adaptive optics (AO) mode, yielding a $\sim1\times1$ arcmin field-of-view sampled by $0.2\times0.2$ arcsec spatial pixels (henceforth spaxels). Data are taken with the blue cut-off filter in place (i.e. the ``nominal'' spectral mode), resulting in wavelength coverage from 4700 to 9350 \AA\, and a spectral sampling of 1.25 \AA\, pixel$^{-1}$. The use of the GALACSI GLAO system roughly doubles the delivered ensquared energy per pixel for MUSE wide-field mode observations, and ensures that all MAGPI targets are observed with an effective seeing of 0.65 arcsec FWHM in $V$-band, or better.

\begin{figure*}
\centering
\includegraphics[width=\textwidth]{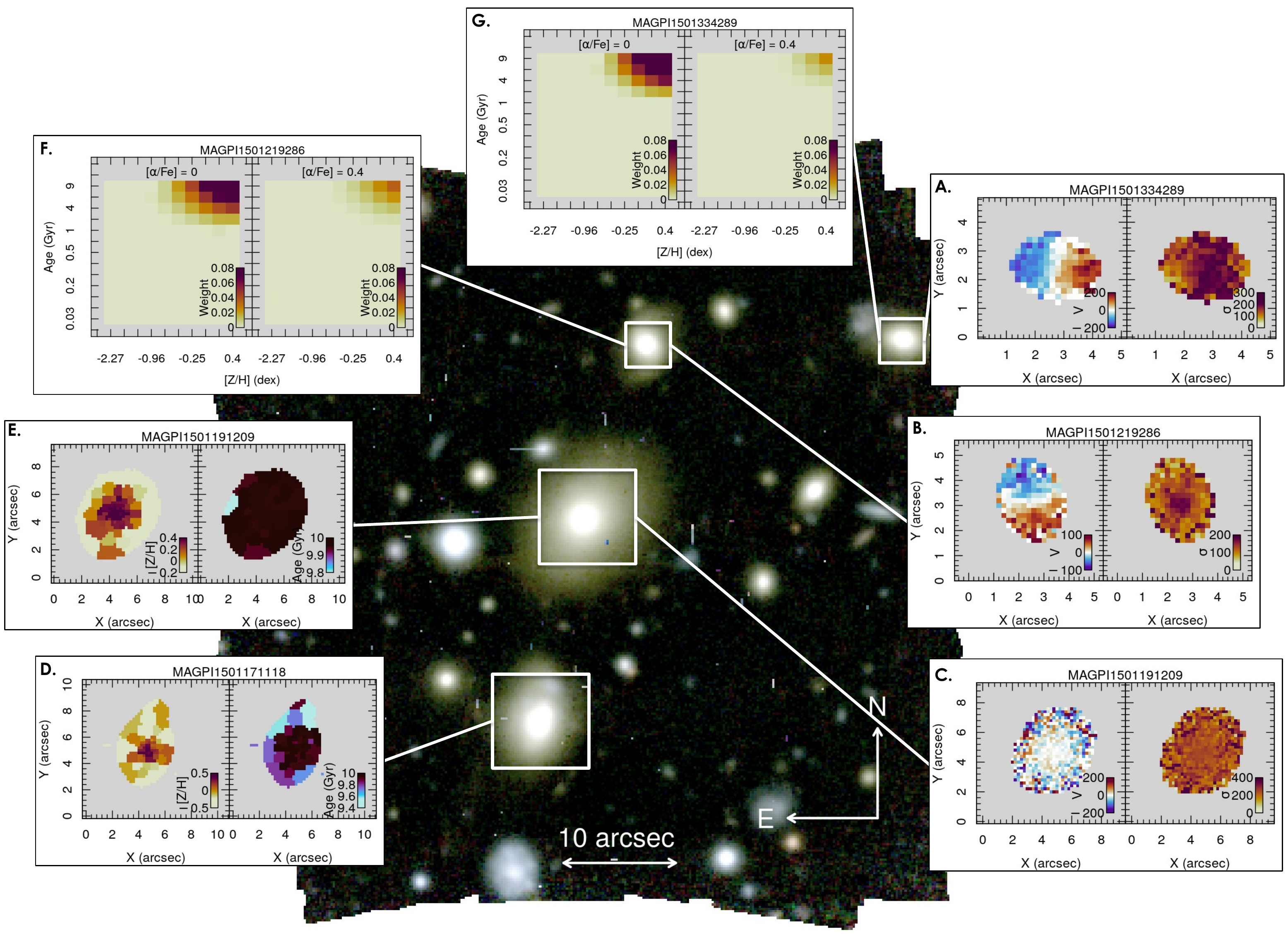}
\caption{Synthetic colour image (${\rm R}=i$, ${\rm G}=r$, ${\rm B}=g_{\rm mod}$) of the MAGPI field G15-J140913. Insets show a variety of high level data products as labelled. Stellar velocity ($V$) and velocity dispersion ($\sigma$) maps are shown for MAGPI1501334289 (Panel A), MAGPI1501219286 (Panel B) and MAGPI15011501191209 (Panel C). Stellar age and metallicity maps are derived for MAGPI1501171118 (Panel D) and MAGPI1501191209 (Panel E), while stellar populations in a 1 arcsec aperture are shown for MAGPI1501219286 (Panel F) and MAGPI1501334289 (Panel G). This figure highlights the exceptional depth and richness of the MAGPI data: our average targets are comparable to the best targets in local IFS surveys.}\label{fig:fieldwithinsets}
\end{figure*}

For each primary target we obtain 6 observing blocks (OBs), comprising $2\times1320$s on source exposures; the total on-source integration time per field is 4.4hr. These long exposures ensure that we reach a S/N of 5 \AA$^{-1}$ per resolution element around 6000-6500 \AA\, in the stellar continuum for individual spaxels at roughly $1\times r_\mathrm{e}$, where the typical surface-brightness for galaxies in our primary sample is $\mu_\mathrm{R}= 23-23.5$ mag arcsec$^{-2}$, and allows us to reliably constrain the first- and second-moments of the line-of-sight velocity distribution \citep[e.g.][]{Bender94,vandeSande17a}. Individual exposures are spatially offset (dithered) and rotated to reduce the impact of the MUSE slicer pattern and/or detector systematics on the final combined frames. The final exposure covers $\sim1.17$ arcmin$^2$ as a result of the adopted dithering and rotation pattern (see, e.g., Fig. \ref{fig:fieldwithinsets}).

\subsection{MAGPI in context} \label{section:comparison}
With a total survey area of $\sim$56 arcmin$^{2}$, 4 hour on-source exposures, and GLAO-corrected image quality (see \S~\ref{section:sample} and \ref{section:observations}), MAGPI fills a niche between the wide-area, shallow MUSE-Wide survey \citep{Herenz17, Urrutia19} and the deeper but narrow-fields MUSE HDFS \citep{Bacon15}, and MUSE HUDF \citep{Bacon17}. The depth and AO-resolution of the MAGPI observing campaign allows further investigation of the evolution of galaxies across cosmic time. Beyond the galaxies selected at $z\sim0.3-0.4$, MAGPI will enable science utilising star-forming galaxies identified through strong optical emission lines ($0<z<1.5$) and Lyman alpha (Ly$\alpha$) emission ($2.9<z<6.0$). The targeting strategy for MAGPI fields can reduce the effects of cosmic variance, for example on the Ly$\alpha$ luminosity function, faced by surveys mainly targeting the deep legacy fields.

Fig. \ref{fig:spatialres} compares the relative spatial resolution of stellar and ionised gas kinematic with lookback time for present and ongoing major IFS campaigns. The science goals of MAGPI are highly complementary to previous and ongoing IFS studies of \emph{ionised gas and stars} in the nearby galaxy population such as e.g. SAURON\citep{Bacon01,deZeeuw02}, DiskMass \citep{Bershady10}, ATLAS$^{\rm 3D}$ \citep{Cappellari11a}, SAMI \citep{Croom12}, TYPHOON \citep{SturchMadore12}, CALIFA \citep{Sanchez12}, MASSIVE \citep{Ma14}, MaNGA \citep{Bundy15}, GHASP \citep{Poggianti17}, Fornax3D \citep{Sarzi18}, MAD \citep{ErrozFerrer19} and Hector \citep{Bryant20}. Despite reaching to nearly twice the lookback time of these existing IFS surveys, MAGPI will deliver a spatial resolution comparable to MaNGA, SAMI, and MASSIVE (Fig. \ref{fig:spatialres}, left panel), facilitating evolutionary studies of massive galaxy kinematics. MAGPI also targets a key epoch between current IFS datasets and future resolved observations at $z>1$ using JWST and ELTs. 

%K-CLASH
With complementary science goals and a sample of 191 star forming galaxies at $0.2 < z < 0.6$, the new IFS survey K-CLASH (K-band Multi-Object Spectrograph Cluster Lensing And Supernova survey with Hubble, \citealt{Tiley20}; \citealt{Vaughan20}) focused on H$\alpha$ emission from ionised gas presents new opportunities for productive scientific synergies with MAGPI.
The right hand panel of Fig. \ref{fig:spatialres} shows how MAGPI strategically links local IFS surveys of the ionised gas to their high-redshift counterpart such as e.g. IMAGES \citep{Yang08}, AMAZE/LSD \citep{Maiolino08}, MASSIV \citep{Contini12}, KMOS$^{\rm 3D}$ \citep{Wisnioski15, Wisnioski19}, KROSS \citep{Magdis16}, KGES \citep{Stott16}, KDS \citep{Turner17} and SINS/zC-SINF \citep{ForsterSchreiber18} at SINS-like spatial resolution.

%The target redshift of MAGPI corresponds to a key epoch during which galaxies experienced a steady decline in their star formation activity.

\subsection{Data reduction}
\label{section:reduction}

Here we briefly outline the relevant data processing steps used to transform the raw MUSE data into flux calibrated and combined cubes for each MAGPI field; a more detailed description of the MAGPI reduction procedure and quality control will be provided in Mendel et al. (in prep.).

First, raw data are processed using {\sc pymusepipe}\footnote{\url{https://github.com/emsellem/pymusepipe}}, which acts as an interface to the ESO MUSE reduction pipeline \citep{Weilbacher12,Weilbacher20}, as well as additional tools for illumination correction and sky subtraction. The main processing steps include bias and overscan subtraction, flat fielding, wavelength calibration, and measurement of the instrumental line-spread-function. Following this initial processing of the science exposures we generate white-light images from the MUSE data, and use these to derive the final output coordinate grid as well as correct for astrometric offsets between the individual cube coordinate systems \citep[due to, for example, ``derotator wobble''][]{Bacon15}. We reconstruct the final cubes and apply a correction for telluric absorption using standard MUSE pipeline tools.

Final processing of the individual MUSE science exposures is performed outside of the standard pipeline using the {\sc CubeFix} (S. Catalupo 2020, in prep.) and \text{Zurich Atmosphere Purge} \citep[ZAP][]{Soto16} packages. We first reconstruct individual exposures onto their final coordinate grid, derived as described above.  We then correct for spatially- and spectrally varying illumination using {\sc CubeFix}, which uses the sky (continuum and lines) as a spatially-uniform reference to re-calibrate individual MUSE slices and IFUs \citep[see][for more details]{Borisova16}. Sky subtraction is then performed using ZAP, which relies on reconstructing the sky in each MUSE $0.2\times0.2$ arcsec spaxel based on a set of principal components derived from the cube itself. The initial illumination-corrected and sky-subtracted cubes are then combined using a 3 $\sigma$ clipped median.  In practice, {\sc CubeFix} and ZAP are applied iteratively, where at each iteration bright sources are masked based on the combined data cube from the previous iteration and {\sc CubeFix} and ZAP are re-run.  In nearly all cases a single subsequent iteration of {\sc CubeFix} and ZAP is sufficient.

\subsubsection{Source detection}

Data products for individual targets are created from the reduced MAGPI cubes. Synthetic white-light, $r$ and $i$-band images for each field are created using the {\sc mpdaf} python package\footnote{\url{https://github.com/musevlt/mpdaf}}. We also create a modified synthetic $g$-band image ($g_{\rm mod}$) because the MUSE nominal wavelength range only partly covers the $g$-band filter range. Then the {\sc ProFound} R package \citep{Robotham18, Robotham18code} is used to detect objects in the white light-image above a threshold of $3\times$ RMS$_{\rm sky}$ and produce a preliminary segmentation map. Similarly to \citet{Bellstedt20}, this segmentation map is then manually adjusted to join mistakenly split segments or remove visibly spurious detections. {\sc ProFound} is used once more to finalise photometric properties using the $r$ and $i$-band images, these include $r_e$ (approximate elliptical semi-major axis containing half the flux), photometric position angle (${\rm PA}_{\rm phot}$), axis ratio and apparent magnitudes, for every object detected in the field. Additional faint emission-line sources are found using custom software with segments added to the full segmentation map.

Unique 10-digit MAGPI IDs are assigned as a concatenation of the 4 digits \texttt{FieldID} (see Table \ref{tab:targetlist2}) and the $3+3$ digits (X,Y) position of the brightest pixel in the white-light image. Objects with an $r$-band $r_e > 0.7$ arcsec FWHM are deemed ``resolved''. For all resolved targets in the field, a series of aperture spectra (0.5, 1, 1.5 and 2 $r_e$ elliptical, as well as 1, 2 and 3 arcsec circular, see examples in Fig. \ref{fig:spectra}) and a ``minicube'' are produced using {\sc mpdaf}, while masking nearby objects based on the segmentation map to avoid contamination. A 1 arcsec aperture spectrum and minicube are also created as above for all unresolved targets in the field. 
We use the {\sc qxp} (Davies et al. in prep.) package in R to measure the redshift, $z_{\rm spec}$, of all objects in the field using these 1 arcsecond aperture spectra.
{\sc qxp} is a modified version of {\sc AutoZ} \citep{Baldry14}, that is currently used for the Deep Extragalactic VIsible Legacy Survey \citep[DEVILS][]{Davies18} and is in development for the core \textit{4-metre Multi-Object Spectroscopic Telescope} (4MOST) L2 redshifting pipeline. Objects with redshift probability values ($p\ge0.98$) are considered secure.

\subsection{Derived quantities}
We present a description of the derived observational quantities shown in this work. The methods described below are under ongoing development and may be improved in subsequent data releases, which will describe relevant changes as required.

%From Francesco:
\subsubsection{Kinematic maps}\label{sec:stellarkin}

Stellar kinematics are extracted spaxel-by-spaxel, but we exclude masked regions, as well as individual spaxels with median $\mathrm{S/N} < 3 \, \mathrm{pixel}^{-1}$. We use the {\sc python} implementation  of the penalised Pixel Fitting program \citep[hereafter {\sc pPXF}][]{CappellariEmsellen04,Cappellari17} and the IndoUS stellar template library \citep{Valdes04}.
The choice of an empirical template library over a synthetic one is motivated by reported discrepancies between synthetic stellar population spectra and observed spectra of local galaxies \citep[][fig.~25]{vandeSande17a} and globular clusters \citep[][figs.~14 and 17]{Conroy18}. 
For the stellar template spectra (hereafter simply: templates), we assume a fixed spectral resolution of $1.35 \, \text{\AA}$ \citep[Gaussian FWHM;][]{Beifiori11}; before fitting, templates are convolved to match the spectral resolution of the MAGPI data as measured from sky lines in the reduced and combined data cube (Mendel et al. in prep.). 

The IndoUS library contains stars with incomplete spectral coverage: we remove 450 templates with gaps in the rest-frame range $\lambda < 7300 \, \text{\AA}$, bringing the number of templates available for the fit to 823. This large number of templates is required to accurately fit high S/N spaxels, but provides excessive freedom for fitting lower S/N ($\mathrm{S/N}\lesssim 15\,\mathrm{pixel}^{-1}$) spaxels, increasing the kinematic uncertainties unnecessarily. To overcome this limitation, we adopt the strategy of the SAMI Galaxy Survey \citep{vandeSande17a}: we pre-select a set of $\approx 15$ templates by fitting the spectrum of a set of elliptical annuli.
These spectra are constructed by adding the spaxels inside an annulus of minimum width equal to one spaxel, and increasing the width until a minimum $S/N = 25 \, \mathrm{pixel}^{-1}$ is reached (or until no more spaxels are available). To fit these spectra, we use the trimmed IndoUS templates, a 12\textsuperscript{th}-order additive Legendre polynomial, a Gaussian line-of-sight velocity distribution (LOSVD), and $c \, z_\mathrm{spec}$ and $\sigma=200 \, \mathrm{km \, s^{-1}}$ as initial guess for the velocity and velocity dispersion, respectively. We mask spectral regions affected by sky emission lines, nebular emission lines and by the AO laser. After the fit, the best-fit spectrum of each annulus is stored. Subsequently, to fit the spectrum of a given spaxel, we first determine a set of intersecting and adjacent annular bins: any annulus intersecting the spaxel, as well as any annulus adjacent to an intersecting annulus. We retrieve the best-fit spectra of each selected annulus, and use this set of spectra as templates for {\sc pPXF}. The same fitting procedure that was used to fit the annular bins is applied to the unbinned spaxels. For both the annular bins and the subsequent fit on individual spaxels, we run pPXF once to estimate the $\chi^2$ per degrees-of-freedom of the fit, then re-scale the input noise spectrum by this value and run pPXF again with the \texttt{clean} keyword. Example resulting stellar kinematic maps are shown in Fig. \ref{fig:fieldwithinsets}.

We measure ionised gas velocity, velocity dispersion, and flux using a set of Gaussian fits to the continuum-subtracted data.  For each spaxel, we first remove the continuum using the best fit stellar kinematics and templates described above.  We then fit the residual spectra using a set of 22 emission lines extending from [OII]$\lambda$3727 to [SII]$\lambda$6732, where the width and relative velocity of all lines are tied. We note that our assumption of a single Gaussian line profile is inaccurate in the presence of multiple kinematic components (e.g. shocks, AGN emission, outflows, etc.); more detailed modelling of the ionized gas kinematics is the subject of future work (Gupta et al. in prep).

%Following couple paragraphs were provided by Sam Vaughan along with the stellar population maps.
\subsubsection{Stellar populations}\label{sec:stellarpopsmaps}
The method for measuring 2D stellar population maps shown in Fig. \ref{fig:fieldwithinsets} will be described in detail in Vaughan et al. (in prep.), but we provide a brief summary here.
First, the minicubes are adaptively binned to an approximately equal S/N ratio of 20 using the Voronoi Tesselation algorithm of \cite{Cappellari03}. We extract flux and variance spectra from each Voronoi bin by summing the appropriate spaxels from the flux and variance cubes in each spectral slice. 

We then use the full spectral fitting code {\sc pPXF} to fit simple stellar population models from the MILES library of \citet{Vazdekis15} to each Voronoi bin. We only include templates that are younger than the age of the Universe at the redshift of our sample, which is $\sim$10 Gyrs. The templates range in metallicity from -2.27 to +0.4 dex, age from 0.03 to 9 Gyrs and can take two values of [$\alpha$/Fe] abundance of 0.00 and +0.4 dex. Each template assumes a Salpeter initial mass function \citep{Salpeter55}.  The normalisation of the templates is set such that the recovered stellar population parameters are mass-weighted, and we use a 10\textsuperscript{th}-order multiplicative Legendre polynomial to correct for large-scale differences in the continuum shape between the templates and observed spectra. 

We also include templates for a number of common emission lines during the fitting procedure, split into two kinematic components. Emission lines in the same kinematic component are constrained to have the same line-of-sight $V$ and $\sigma$. The first component contains a series of emission lines corresponding to the Balmer series (H$\alpha$ to H$\theta$). The flux values of each line in the Balmer series are fixed according to the intrinsic Balmer decrement for Case B recombination with electron temperature $T=10^4$~K and a number density of $n=100$cm$^{-3}$ \citep{Dopita03}, with their fluxes scaled up or down in lockstep (i.e. using the \texttt{tie\_balmer} keyword in {\sc pPXF}). We also fit for reddening from these Balmer lines using a Calzetti extinction curve \citep{Calzetti00}. 
The second component corresponds to the [OIII]$\lambda4959,\lambda5007$ doublet; the [OI]$\lambda6300,\lambda6364$ doublet; the [NII]$\lambda6548,\lambda6583$ doublet; and the [SII]$\lambda6716,\lambda6731$ doublet. In each case, we use the \texttt{limit\_doublets} keyword in {\sc pPXF} to limit the fluxes of each doublet component to be between the values allowed by atomic physics. 

As was done with the stellar kinematics, we run {\sc pPXF} once to estimate the $\chi^2$ per degrees-of-freedom of the fit, then re-scale the input noise spectrum by this value and run {\sc pPXF} again with the \texttt{clean} keyword. This iteratively clips the spectrum of outliers and bad pixels \citep[see][for further details]{Cappellari17}. As we are not interpreting the weights on individual templates for the stellar population maps, we do not use regularisation (i.e. \texttt{regul} = 0) for this step. After the fitting, we extract the weighted average age and metallicity of each Voronoi bin by summing over the best-fitting weights from {\sc pPXF}. Our results correspond to the mass-weighted average quantity for each spectrum.

%Tania's blurb
%\subsubsection{Integrated stellar population measurements}\label{sec:stellarpops}
For galaxies with limited spatial extent, we are able to measure global ages and metallicities from the integrated spectra. The integrated stellar population parameters are measured following the same method to the spatially resolved maps, but with a regularisation value of \texttt{regul} = 100. %The main difference is the handling of emission lines; for the integrated spectra the emission lines are masked rather than fit. The process of masking emission lines also masks unwanted features, including residual telluric lines, bad pixels, and noisy regions, which is more important in low S/N spectra. As with the spatially resolved maps we measure both luminosity and mass weighted parameters, with luminosity-weighted parameters generally having lower uncertainties. See \citet{Barone20} for a full description of the method used for the integrated stellar population fits, as well as a presentation of the typical uncertainties based on S/N. 
We demonstrate the integrated stellar population fits in Fig. \ref{fig:fieldwithinsets} by showing the template weights for galaxies MAGPI1501334289 and MAGPI1501219286.

\subsection{Theoretical dataset} 
\label{section:datasims}
%Nitty gritty details of the simulations and perhaps how the dataset will be produced.
%One of the key aims of MAGPI is to connect observations with simulations.
%As described in \S~\ref{sec:theorycase}, the MAGPI theory library includes {\sc EAGLE} \citep{Schaye15,Crain15}, {\sc Magneticum} \citep{Teklu15,Schulze18}, {\sc HORIZON-AGN} \citep{Dubois16}, {\sc Illustris-TNG100} \citep{Pillepich18, Naiman18, Springel18, Nelson19b} and TK15 \citep{Taylor15,Taylor17}.
This section summarises relevant differences between the simulations in our library and outlines planned theoretical MAGPI data products. A more detailed overview of the simulations in the MAGPI theoretical library can be found in Appendix \ref{sec:simsdescription}.

\subsubsection{Simulations}\label{sec:simsmainbody}

Broadly all the simulations include the same key physical processes: metal cooling, photo-ionisation, star formation, stellar evolution and chemical enrichment, feedback from stars and supermassive black holes. The key differences reside in how these processes are modelled in detail (see \citealt{Vogelsberger20} for a recent review). Table~\ref{tab:simsdetails} shows key technical information about the cosmological hydrodynamical simulations currently in our suite. 
We show both the highest spatial resolution achieved for both gas and dark matter, however, we caution that for galaxy structure and kinematics what matters is the spatial resolution of the dark matter (rather than the gas or stars; \citealt{Ludlow20}).

\begin{table*}
    \centering
    \begin{tabular}{c|c|c|c}
         \hline
         Simulation & Volume & Particle mass gas/DM & Spatial resolution \\
         \hline
         {\sc EAGLE}& $100^3$ & $1.8\times 10^6$/$9.7\times 10^6$ & 0.7/0.7 \\  
         {\sc Magneticum} & $68^3$ & $1.0\times 10^7$/$5.1\times 10^7$ & 1.99/1.99 \\
         {\sc HORIZON-AGN} & $142^3$ & $10^7$/$8\times 10^7$ & 1/1\\
         {\sc Illustris-TNG100} & $111^3$ & $1.4\times 10^6$/$7.5\times 10^6$ & 0.19/0.74\\
         {\sc TK15} &  $35.7^3$ & $1.4\times 10^7$/$7\times 10^7$ & 1.6/3.2\\
         \hline
    \end{tabular}
    \caption{Key information of the simulations currently part of the MAGPI theory library. For each of these we show the simulated cosmological volume (in units of comoving $\rm Mpc^3$), initial gas and dark matter particle masses (in units of $\rm M_{\odot}$) and the highest spatial resolution for gas and dark matter (in units of comoving $\rm kpc$). The Magneticum simulation employs a smaller softening for stellar particles, corresponding to a spatial resolution of 1 kpc.}
    \label{tab:simsdetails}
\end{table*}

The hydrodynamic techniques used by the simulations in Table~\ref{tab:simsdetails} are varied, with {\sc EAGLE}, {\sc Magneticum} and TK15 employing smooth particle hydrodynamics, {\sc HORIZON-AGN} employing Adaptive Mesh Refinement and {\sc Illustris-TNG100} an unstructured mesh strategy. These simulations also adopt different cosmological parameters: {\sc EAGLE} adopts \citet{Planck14}, {\sc Magneticum} and {\sc HORIZON-AGN} adopt \citet{Komatsu11}, {\sc Illustris-TNG100} adopts \citet{Planck16} and TK15 adopts \citet{Hinshaw13}. Despite these differences in hydrodynamics solver and cosmology, most of the differences in the predicted properties of the galaxy population are due to the modelling of physical processes that happen below the spatial scales typically resolved.

Briefly, {\sc EAGLE}, {\sc HORIZON-AGN} and TK15 use thermal energy injection to model stellar feedback, while {\sc Magneticum} and {\sc Illustris-TNG100} increase the velocity of nearby particles and decouple them from the hydrodynamic calculation for a period of time. {\sc Magneticum}, {\sc HORIZON-AGN} and {\sc Illustris-TNG100} implement AGN feedback so that there are two modes that are distinct for black holes accreting close to the Eddington limit and those well below; meanwhile {\sc EAGLE} and TK15 model AGN feedback as a single mode of energy injection. A more detailed description of each of these simulations is presented in Appendix \ref{sec:simsdescription}.
%CONSIDER REINCORPORATING?: A final important difference between these three simulations is the simulated cosmological volume, with {\sc HORIZON-AGN}, {\sc EAGLE} and {\sc Magneticum} simulating volumes of $(142 \rm \,Mpc)^3$, $(100\,\rm Mpc)^3$ and $(59\,\rm Mpc)^3$, respectively. The latter is responsible for the lower number statistics evident in {\sc Magneticum} compared to {\sc HORIZON-AGN} and {\sc EAGLE}. 
%The \citet{Taylor17} simulation offers a comparatively smaller cosmological volume still.

\begin{figure}
        \begin{center}
                \includegraphics[trim=0mm 0mm 0mm 0mm, clip,width=0.5\textwidth]{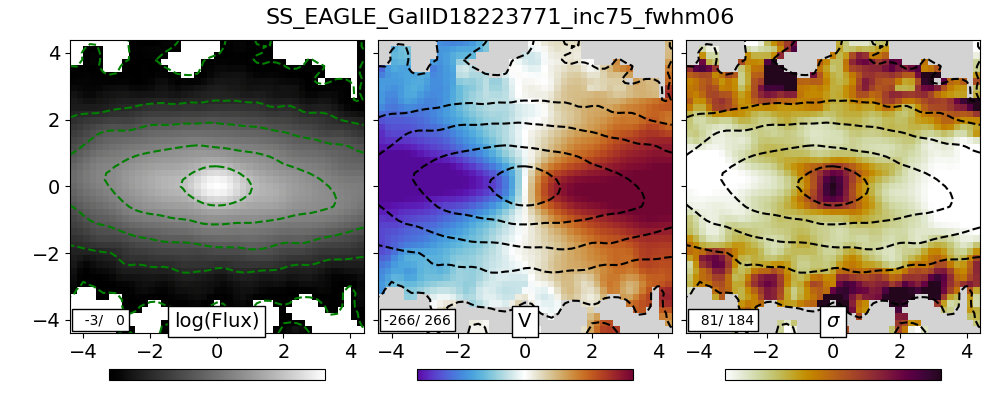}
                \includegraphics[trim=0mm 0mm 0mm 0mm,clip,width=0.5\textwidth]{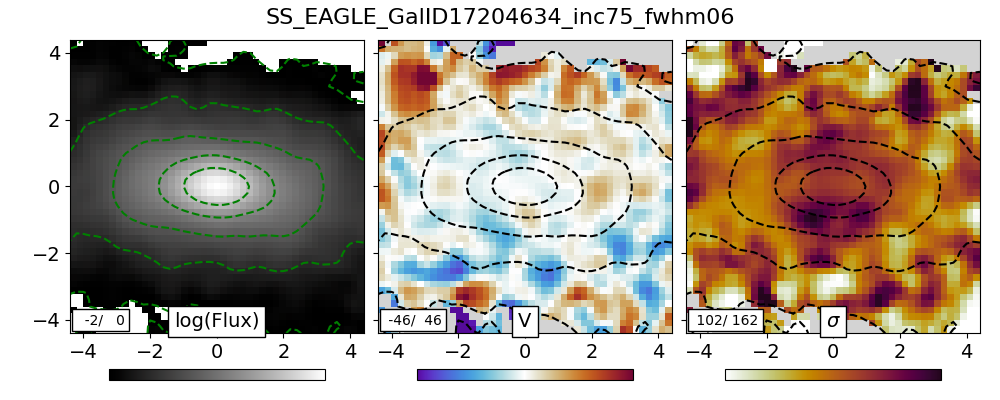}
                \includegraphics[trim=0mm 0mm 0mm 0mm,clip,width=0.5\textwidth]{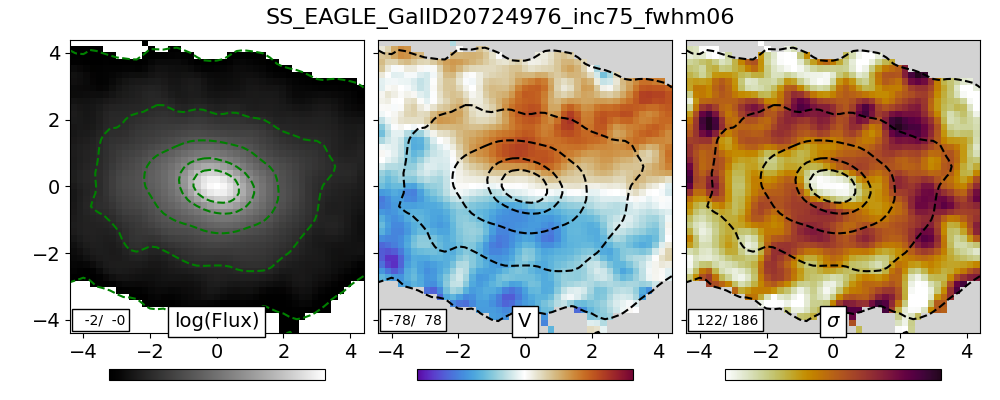}
                 \includegraphics[trim=0mm 0mm 0mm 0mm,clip,width=0.5\textwidth]{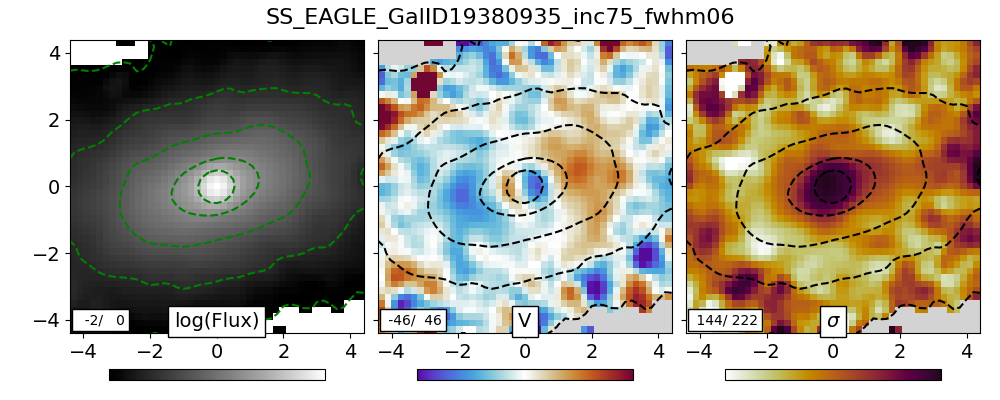}
                  \includegraphics[trim=0mm 0mm 0mm 0mm,clip,width=0.5\textwidth]{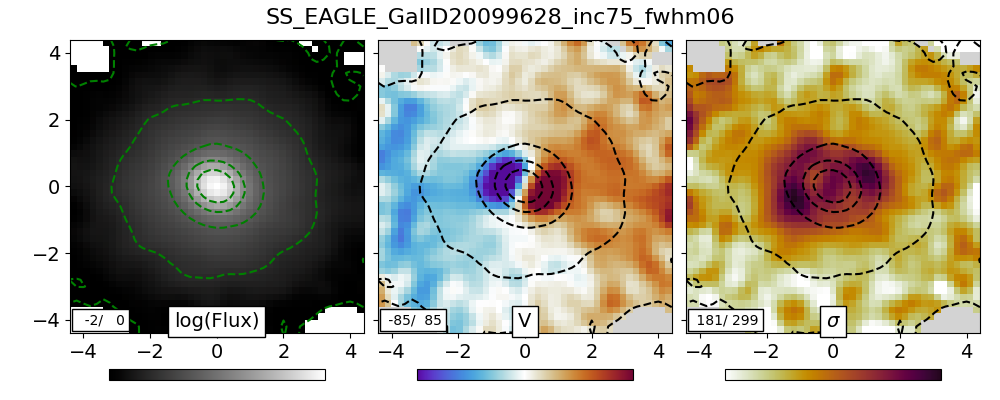}
                \caption{Examples of MAGPI-like maps produced using {\sc EAGLE} galaxies and the post-processing software {\sc SimSpin}, using the specifications of MAGPI. These maps show the quality of maps we expect for MAGPI and the diversity of kinematic classes we expect. Ticks in the x- and y-axes refer to kpc. From left to right the images show flux, line-of-sight velocity and velocity dispersion maps. From top to bottom we show example maps of a typical fast rotator, a slow rotator, a prolate galaxy, a galaxy with a kinematically-decoupled core and a 2$\sigma$ galaxy at $z\approx 0.3$ in {\sc EAGLE}. The range in colours is shown at the bottom of each panel. Each galaxy has been inclined to 75 degrees and we adopt a FWHM of $0.6$ arcsec. The simulation's GalaxyID (which can be used to cross-correlate with the public {\sc EAGLE} database;  \citealt{McAlpine16}) is labelled for each row of panels.} 
                \label{fig:examplemaps}
        \end{center}
\end{figure}

\subsubsection{Data products}

To fulfil our goal of making full use of the simulation suite, we present datasets in two ways for the simulations that are currently in our theory library. The first one consists of making relevant measurements within the simulations in a consistent manner directly comparable to MAGPI observational data. The second one consists in creating 3D cubes of galaxies that can be analysed with the same tools we use for the observations. Below we provide a short description of these two approaches:
\begin{enumerate}
\item \textit{Providing tabulated predictions computed in a consistent manner.} We follow the strategy of \citet{vandeSande19} and ask team members with access to and expertise with the different simulations to provide measurements of a range of physical properties of galaxies that science projects are aiming at using. Currently these include: stellar mass, star formation rate, halo mass, central/satellite distinction and r-band $r_{\rm e}$. Several properties are then computed within integers of $r_{\rm e}$ (1 and 2): specific stellar angular momentum ($j^*$), stellar spin parameter ($\lambda_r$), star-forming gas metallicity ($\log(\rm O/H) + 12$) and radial metallicity slope of the star-forming gas ($\alpha_r$). These properties are provided at several redshifts between $z=0$ and $z=1$, but most critically at $z\approx 0.3$, which is the redshift of interest for MAGPI. This enables the analysis of evolutionary trends that we can then connect with existing $z\approx 0$ and $z\gtrsim 0.5$ surveys.
\item \textit{Creating synthetic cubes of galaxies in the simulation suite.} We create data cubes matched to the MAGPI observations: a spatial pixel of $0.2$~arcsec, a velocity pixel of $1.25$\AA, a line-spread function (LSF) of FWHM$=2.63$\AA, and observational ``noise'' using {\sc SimSpin}\footnote{\url{https://github.com/kateharborne/SimSpin}} \citep{Harborne20}. These MAGPI mock cubes use galaxies at the redshift of MAGPI and are projected to a redshift distance of $z=0.25-0.35$ to match the observation specifications. These cubes keep the number of pixels within $r_e$ approximately fixed. So far these have been created for stars only with the purpose of studying stellar kinematics. In the future they will be extended to include stellar populations and gas properties as well. Examples of the existing cubes are presented in Fig.~\ref{fig:examplemaps}, visualized using {\sc Pynmap}\footnote{\url{https://github.com/emsellem/pynmap}}. These cubes are provided in FITS format to facilitate their analysis using the same tools as used by observers in the team, and are generated at $4$ different inclinations ($30$, $45$, $60$ and $75$ degrees) with the aim of investigating the systematic effect this can have.
\end{enumerate}

\section{Early Results}\label{section:results}

%\begin{figure*}
%\includegraphics[width=17cm]{Figures/redshiftmaps.pdf}
%\caption{\textcolor{red}{Move figure to Trevor's paper?} Inverted synthetic $r$-band image of the MAGPI field G15-J140913 with coloured output from {\sc FourXP}. The redshift distribution for all reliably fit spaxels (left) and for those within a narrow redshift range centered on the central galaxy's redshift (middle) are shown. The right panel shows the type of template favoured by {\sc FourXP} (S: star, P: passive galaxy, SF: star forming galaxy, QSO: quasar). The left panel also shows objects that were detected by {\sc ProFound}, but without a reliable {\sc FourXP} redshift, as black crosses. For both the left and right panels, values obtained from synthetic 1 arcsecond aperture spectra are shown as filled circles. There is a rich diversity of galaxy types and dynamics as well as potentially interesting background and foreground objects in this field alone. In all panels, North is up and East is left.}\label{fig:zmap}
%\end{figure*}

\begin{figure*}
\includegraphics[width=\textwidth]{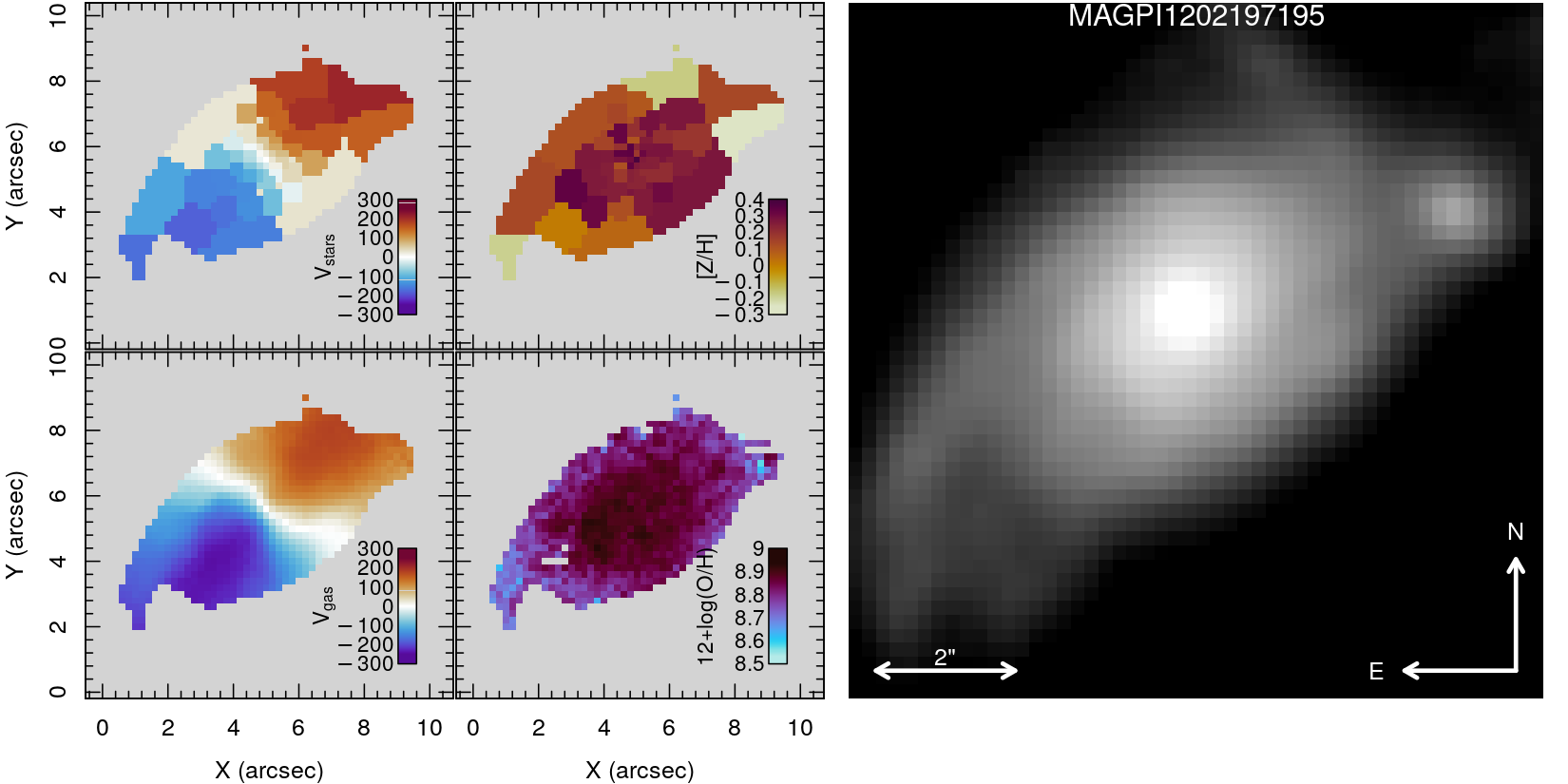}
\caption{Example primary MAGPI galaxy with significant ionised gas component allowing for direct comparison of stellar and gas properties at $z\sim0.3$. Right: Synthetic white-light image of MAGPI1202197195, the central galaxy for field G12-J114121. A 2 arcsec scale ($\sim8.9$~kpc) is shown for reference. Left panels show the star (top) and gas (bottom) kinematic maps, while middle panels show stellar (top) and gas-phase (bottom) metallicities.
North is up and East is left.}\label{fig:gasandstars}
\end{figure*}

We demonstrate selected aspects of the observational and simulated data to date.

\subsection{Observations}
At the time of writing, MAGPI observations are under way with data available for 15 completed fields (G12-J113850, G12-J114121, G12-J114123, G12-J114238, G12-J115219, G12-J120038, G12-J120759, G12-J121953, G12-J122223, G15-J140913, G15-J142228, G15-J142332, G15-J143616, G15-J143809 and G23-J223757), 8 partly observed (G15-J141428, G15-J143840, G15-J145221, G23-J224045, G23-J224634, G23-J230506, G23-J231312 and G23-J231911) and 2 archive fields (Abell 370 and Abell 2477; see Table \ref{tab:targetlist2} and Figs. \ref{fig:fieldsgallery} and \ref{fig:archivefields}).

In what follows, we present selected observational data for two MAGPI fields, the first has intermediate density ($M_{\rm halo}/M_{\odot}=13.16$): G15-J140913 (\texttt{FieldID} $=1501$). There are 19 `resolved' ($r$-band $r_e>0.7$ arcsec) galaxies at the redshift of interest in this field. 
%\textcolor{red}{Move rest of this paragraph to Trevor's paper? Fig. \ref{fig:zmap} shows the output of the {\sc FourXP} redshifting procedure for both the aperture spectra and individual spaxels with reliable redshifts ($p>0.98$). As well as satellites of the passive central galaxy (GAMAID 237785 or MAGPI1501191209), {\sc FourXP} identifies several passive and star-forming satellites, as well as foreground stars and background galaxies. As well as a simple spectral classification for the objects in the field, this method also returns crude velocity maps for some of the resolved targets in the field.}

Fig. \ref{fig:fieldwithinsets} shows the synthetic $g_{\rm mod}ri$ image of the G15-J140913 field with selected stellar kinematic and populations as insets. The methodology employed to derive the results presented in Fig. \ref{fig:fieldwithinsets} insets are described in \S~\ref{sec:stellarkin} and \S~\ref{sec:stellarpopsmaps}. Fig.~\ref{fig:fieldwithinsets} shows that the central galaxy (GAMAID 237785 or MAGPI1501191209) is a clear slow rotator, has a negative stellar metallicity gradient and a uniformly old stellar ages. Neighbouring galaxies MAGPI1501334289 and MAGPI1501219286 show clear rotation and a central peak in their velocity dispersion maps. 
Nearby galaxy MAGPI1501171118 exhibits a negative stellar metallicity gradient and hints of a negative age gradient. Detailed stellar populations for two other galaxies in the field (MAGPI1501219286 and MAGPI1501334289) suggest a prominence of old and metal rich stars with predominantly solar [$\alpha$/Fe] abundances in both galaxies.

In addition to stellar population maps and star formation histories as shown above, systems that contain significant ionised gas (e.g. Fig. \ref{fig:gasandstars}) also enable the study and comparison of the gas-phase metallicities and dynamics. In MAGPI1202197195, the bright central in field G12-J114121 (\texttt{FieldID} $=1202$, $M_{\rm halo}/M_{\odot}=14.78$, see Table \ref{tab:targetlist2}), extended maps of both the stellar and ionised gas components can be derived. The kinematic maps enable e.g. the computation of $\lambda_r$, kinematic offset between gas and stars, and kinematic asymmetries using kinemetry of both the gas and the stars. The metallicity maps of the gas and stars enable the measurement of metallicity gradients. Gas-phase metallicity gradients represent a key measurable where theoretical models as well as simulations show tension in galaxies at $z\sim 0.3$ (see \S~\ref{section:resultsims}).

We leave the detailed analyses of the observed dynamics, stellar populations and ionised gas properties of MAGPI galaxies to future papers. The early results shown here demonstrate that the MAGPI data are of the anticipated quality and depth to accomplish the survey science goals presented in \S~\ref{section:sciencegoals} and that these observations can be straightforwardly compared with their simulated counterparts produced by the survey theory working group (see \S~\ref{section:resultsims}).

\subsection{Theoretical predictions}
\label{section:resultsims}

\begin{figure}
\includegraphics[trim=7mm 10mm 12mm 14mm, clip,width=0.45\textwidth]{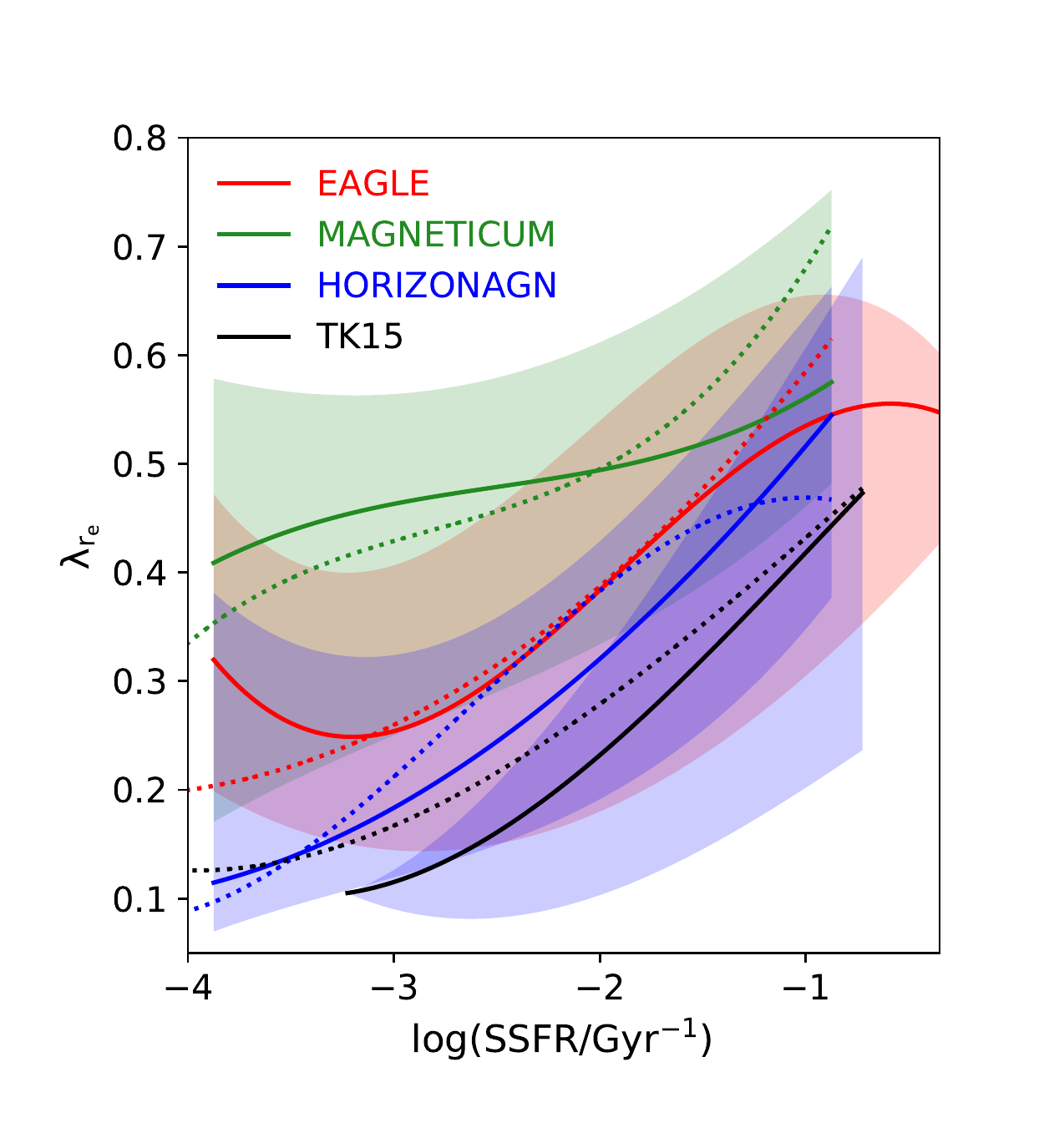}
\caption{Predicted dependence of $\lambda_{\rm r_e}$ on the specific star formation rate for the ``well-resolved MAGPI-like" samples in {\sc EAGLE} (red), {\sc Magneticum} (green) and {\sc HORIZON-AGN} (blue),  and for a ``primary-MAGPI like'' sample in TK15 (black) at $z\approx 0.3$ (solid lines). Solid lines and shaded regions show the smoothed medians and $1\sigma$ percentile ranges, respectively. For reference, we also show the predicted median relation at $z=0$ as dotted lines. All simulations predict $\lambda_{\rm r_e}$ to correlate with the specific star formation rate, but the exact dependence is model-dependent.}\label{fig:lambdaRAtMAGPI}
\end{figure}

Below we explore the theoretical expectation for the dynamical state and metallicity profiles of galaxies at the redshift of MAGPI using a range of galaxy simulation models (see \S~\ref{section:datasims}). 
The main goal of this section is to understand what MAGPI could constrain and measure. To remedy the fact that the simulations introduced in $\S$\ref{sec:simsmainbody} have different cosmological volumes, we decide to randomly sample the simulations to obtain a similar number of galaxies to those we expect for MAGPI. We are interested in two populations of galaxies: 
the ``primary targets'' - galaxies with stellar masses $\ge 10^{10.8}\,\rm M_{\odot}$, of which we expect $\approx 60$; and the ``secondary targets'' - galaxies in the field-of-view of the primary (aka ``satellite'' galaxies) that are expected to be well resolved, i.e. stellar masses $\ge 10^{10}\,\rm M_{\odot}$, of which we expect $\approx 100$. Put together, we refer to this sample as the ``well-resolved MAGPI-like'' sample.

\subsubsection{Dynamical evolution and environmental effects}\label{subsec_simsresultskinematics}
Theoretical stellar kinematic measurements (in this case $\lambda_{\rm r_e}$) are performed as in \citet{vandeSande19} for {\sc EAGLE} and {\sc Magneticum},  and as in \citet{ChoiYi17} for {\sc HORIZON-AGN}. For the TK15 simulation, light-weighted, line-of-sight velocity and velocity square maps were generated. From these, $\lambda_{\rm r_e}$ was computed within a circular aperture of radius $r_{\rm e}$. \S~\ref{section:datasims} presents a description of the physics included in these simulations. An important caveat is that {\sc EAGLE}, {\sc Magneticum}, and {\sc HORIZON-AGN} are sufficiently large as to allow the construction of a well-resolved MAGPI-like sample. This is not the case for the TK15 simulation, which is $\approx 7-63\times$ smaller than the other simulations. Because of this,  we were only able to build a sample from this simulations that resembles the primary targets of MAGPI (i.e. $60$ galaxies with $M_{\star}\ge 10^{10.8}\,\rm M_{\odot}$. Hence, compared to the other simulations, the TK15 sample will be biased towards higher stellar masses. We only include the TK15 simulation when we analyse the dependence of $\lambda_{\rm r_e}$ on the specific star formation rate of galaxies.
%An important difference here is the simulated cosmological volume, which is smaller in {\sc Magneticum}, causing the evident lower number statistics compared to the other two simulations.

%We confirm that the progenitors of today's dispersion-supported slow-rotators, and rotation-supported fast-rotators had indistinguishable kinematics at $z\gtrsim1$ and diverged dramatically towards $z=0$ \citep{Penoyre17, ChoiYi17, Lagos17}; the timescale of the transition is slower in low halo mass environments such that the $\lambda_{\rm r_e}$ distributions diverge maximally with halo mass at $z\sim0.3$ \citep[Fig. \ref{fig:lambdaRevol}]{Lagos18b}. 
The left panels of Fig.~\ref{fig:lambdaRevol} show the 2-dimensional distribution of galaxies in the lookback time vs. $\lambda_{\rm r_e}$ plane for the primary MAGPI-like samples, as defined above, in {\sc EAGLE}, {\sc Magneticum} and {\sc HORIZON-AGN} hydrodynamical simulations.

Fig.~\ref{fig:lambdaRevol} further shows that {\sc EAGLE} and {\sc Magneticum} predict massive galaxies to have a relatively narrow distribution of $\lambda_{\rm r_e}$ at $z=0$ with peaks at $\lambda_{\rm r_e}\approx 0.22$ ({\sc EAGLE}) and $\lambda_{\rm r_e}\approx 0.18$ ({\sc Magneticum}). Contrary to this, {\sc HORIZON-AGN} predicts a broad $\lambda_{\rm r_e}$ distribution at $z=0$ with two  peaks at $\lambda_{\rm r_e}\approx 0.15$ and $\lambda_{\rm r_e}\approx 0.55$. At $z=1$, {\sc EAGLE} and {\sc HORIZON-AGN} predict a peak at $\lambda_{\rm r_e}\approx 0.7$ and $\lambda_{\rm r_e}\approx 0.6$, respectively, while {\sc Magneticum} predicts a peak at a lower $\lambda_{\rm r_e}\approx 0.45$. 
Significant kinematic transformation is seen in all simulations for massive galaxies, but at different cosmic epochs. In {\sc EAGLE}, this happens at $0.3\lesssim z\lesssim 0.6$, in {\sc Magneticum} at $0.5\lesssim z\lesssim 0.8$, while {\sc HORIZON-AGN} predicts most of the transformation to happen at higher redshift, $z\gtrsim 0.8$. 
%Both simulations indicate that the dynamical transformation happens at $z<1$. 
To assess the effect of environment, we compare the distribution of $\lambda_{\rm r_e}$ of the ``primary MAGPI-like'' galaxies in the bottom and top $33^{\rm rd}$ percentiles of the halo mass distribution in the three simulations and refer to those as low and high density environments, respectively. 
As each simulation predicts a different stellar-to-halo mass relation, and the selection of the sample was done in stellar mass, the exact halo mass thresholds defining these percentiles vary between simulations. For {\sc EAGLE} and {\sc Magneticum}, these halo mass thresholds are $\approx 10^{12.9}\, M_{\odot}$ and $\approx 10^{13.4}\,\rm M_{\odot}$, respectively, while for {\sc HORIZON-AGN} these are $\approx 10^{12.3}\,\rm M_{\odot}$ and $\approx 10^{12.9}\,\rm M_{\odot}$.
At the redshift of MAGPI, the three simulations predict different degrees of environmental impact, with {\sc EAGLE} and {\sc HORIZON-AGN} predicting a $\lambda_{\rm r_e}$ distribution skewed to high values at low densities compared to galaxies of the same stellar mass in high density environments. These trends are strong enough that we expect MAGPI to detect them with the primary sample of $60$ galaxies. Although {\sc Magneticum} predicts an environmental effect, the high- and low-density distributions are less distinct to the point that there would not be enough galaxies in the primary MAGPI sample to detect this environmental impact. The fact that these trends arise clearly in two out of the three simulations, after we sample them to have the same expected number of primary targets as MAGPI, provides evidence to state that the survey is designed to have enough massive galaxies to robustly measure their $\lambda_{\rm r_e}$ distribution in high and low-density environments. We thus expect MAGPI to be able to distinguish between these different predictions.

\citet{Wang20} show that in addition to the dependence of $\lambda_{r}$ on $M_*$, $\lambda_{\rm r_e}$ strongly depends on the star formation rate.
In fact, part of the predicted environmental dependence of $\lambda_{\rm r_e}$ in the different simulations comes from how they predict this quantity to vary with star formation activity and stellar mass in galaxies. Fig.~\ref{fig:lambdaRAtMAGPI} shows the expected dependence of $\lambda_{\rm r_e}$ on the specific star formation rate at $z\approx 0.3$ in {\sc EAGLE}, {\sc Magneticum}, {\sc HORIZON-AGN} and TK15 (solid lines).
{\sc EAGLE} and {\sc Magneticum} define star formation rates as instantaneous, while in {\sc HORIZON-AGN} and TK15 this is the average SFR over the past $100$~Myr and $10$~Myr, respectively. Because most galaxies have smoothly declining star formation rates on those timescales, this difference in the way they are measured does not play an important role here (we tested different timescales from $10-100$~Myr and obtained only small differences that do not change the interpretation). Passive galaxies, which are preferentially found in high-density environments, are expected to have lower 
$\lambda_{\rm r_e}$ in the three simulations. However, the exact dependence of $\lambda_{\rm r_e}$ on the specific star formation rate depends on the simulation. {\sc HORIZON-AGN} and TK15 predict the steepest relation, followed by {\sc EAGLE}, while {\sc Magneticum} predicts a shallower dependence. In the four simulations we find that the scatter of the relation is correlated with stellar mass, with lower (higher) stellar masses scattering up (down). In practice this could be tested by comparing where the primary vs. the secondary MAGPI targets lie in this plane, as the former will on average be more massive than the latter. TK15 predicts the lowest $\lambda_{\rm r_e}$ of the four simulations at fixed specific star formation rate. Part of this is due to the fact that from this simulation we were only able to construct a primary MAGPI-like sample rather than the full well-resolved MAGPI-like sample, and as explained above, there is an underlying dependence on stellar mass, where more massive galaxies tend to have lower $\lambda_{\rm r_e}$.
For reference, we also show the $z=0$ predicted relation in Fig.~\ref{fig:lambdaRAtMAGPI} and find that in general all simulations predict that the relation between $\lambda_{\rm r_e}$ and specific star formation rate becomes steeper from $z=0.3$ to $z=0$, except for TK15, which predicts a shallower $z=0$ relation.
%There are also interesting differences as to how $\lambda_{\rm r_e}$ depends on the specific star formation rate for primary and secondary-like targets, which are rooted in how $\lambda_{\rm r_e}$ is predicted to depend on stellar mass in these simulations (with the former being more massive than the latter).} 
%that lines of constant $\lambda_{\rm r_e}$ follow the main sequence closely, while in {they appear steeper. For Magneticum $\lambda_r$ appears to be more strongly correlated with $M_*$ than to SFR as compared to the other simulations. 
MAGPI will be key to unveil the true shape of this relation at intermediate redshifts and hence place fundamental constraints on galaxy formation simulations. Disentangling how kinematic transformation and quenching happen in galaxies and whether these two processes correlate are key questions MAGPI, together with low redshift surveys, can shed light on.

Fig.~\ref{fig:examplemaps} shows kinematic maps of {\sc EAGLE} galaxies built using {\sc SimSpin} \citep{Harborne20} and adopting the specifications of MAGPI (see \S~\ref{section:datasims}).
These maps were selected to display the kinematic diversity expected for MAGPI galaxies, including slow- and fast-rotators, major-axis (prolate) rotation, kinematically decoupled cores and counter-rotating discs (i.e. $2\sigma$, \citealt{Krajnovic11}).
This will offer important constraints on how the angular momentum of stars and ionised gas are correlated and whether important differences are seen with respect to the local Universe.

\subsubsection{Metallicity gradients and the effect of environment, mass and star formation rate}

Gas metallicity gradients, $\alpha_{\rm r_e}$, were measured by selecting gas cells or particles that are actively involved in star formation (and hence are a good proxy for the ionised gas we expect to measure with MAGPI) within a spherical aperture of radius $r_{\rm e}$. We then bin the gas metallicity in equidistant logarithmic radial bins, and measure the slope of the function $\rm log(Z_{\rm gas}/Z_{\odot}) = \alpha_{\rm r_e}\,r + a_{0}$. Here,  $\alpha_{\rm r_e}$ has units of $\rm dex/kpc$, $r$ is in kpc, and $a_0$ is the intersect of the radial profile at $r=0$. 
%\textcolor{red}{Claudia: what is $a_0$?}
%We measure the slope of the radial metallicity profile of gas cells or particles that are actively involved in star formation (and hence are a good proxy for the ionised gas we expect to measure with MAGPI), using all particle/cells at $r<r_{\rm 50}$, with $r_{\rm 50}$ being the r-band half-light radius in kpc. We binned the gas metallicity in ${\rm log$ in equally distant radial bins, and fit it with a linear relation that yields a slope in units of $\rm dex/kpc$. This is done in a consistent manner in the four simulations. We will refer to this slope as $\alpha_{\rm r_{50}}$.

\begin{figure}
\includegraphics[trim=17mm 9.5mm 6mm 7mm, clip,width=0.245\textwidth]{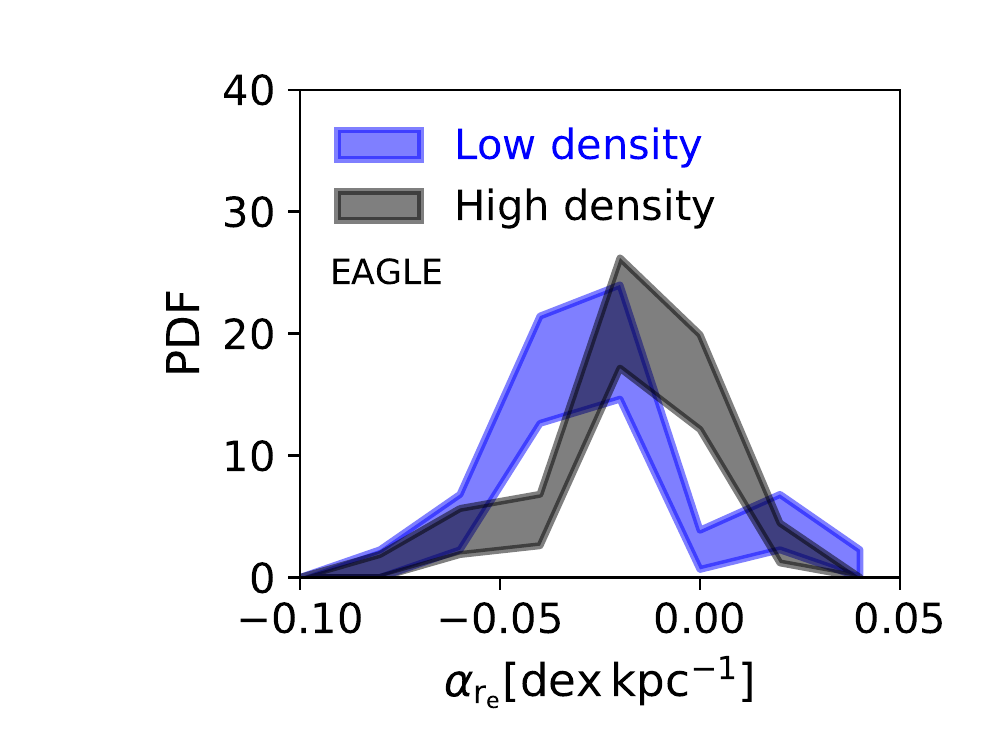}
\includegraphics[trim=22.5mm 9.5mm 6mm 7mm, clip,width=0.228\textwidth]{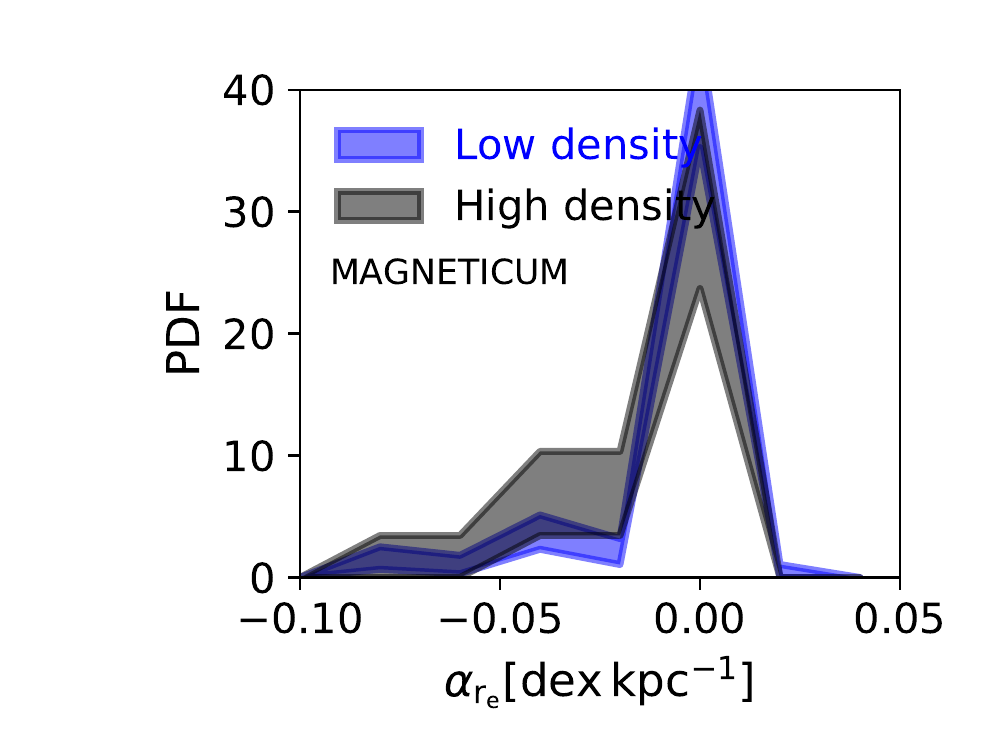}
\includegraphics[trim=17mm 4mm 6mm 7mm, clip,width=0.245\textwidth]{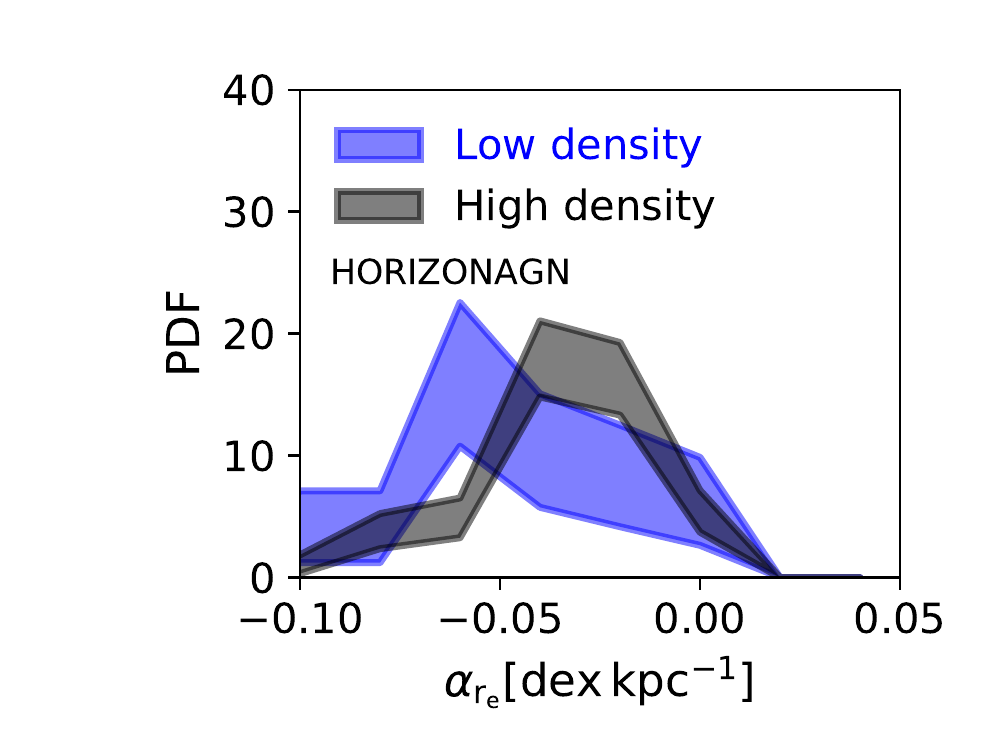}
\includegraphics[trim=22.5mm 4mm 6mm 7mm, clip,width=0.228\textwidth]{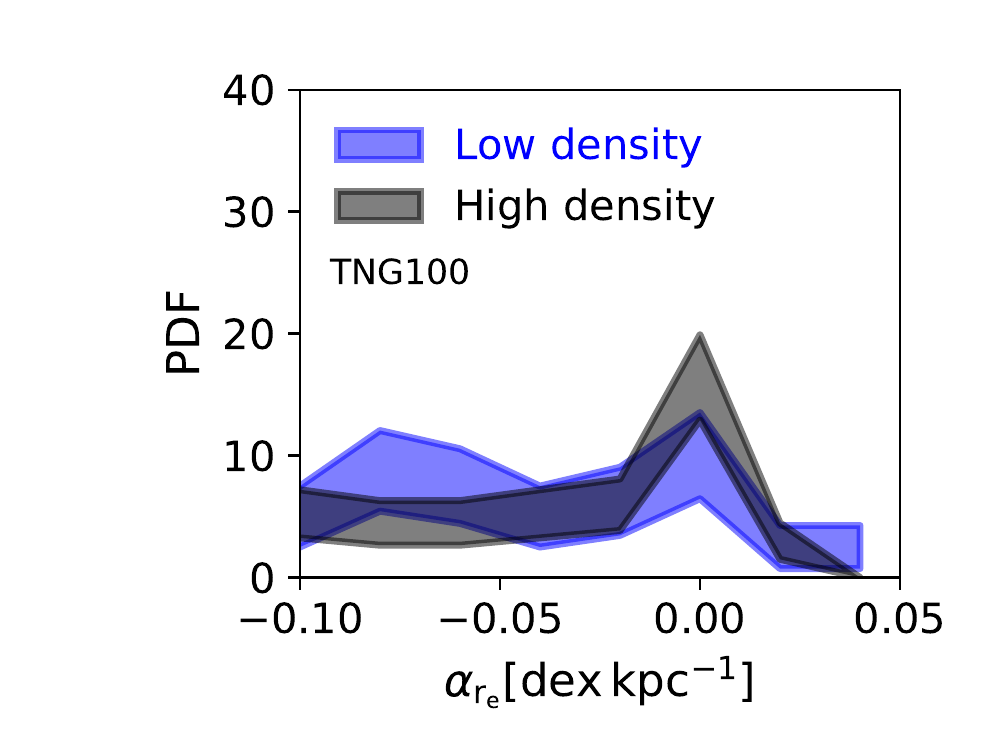}
\caption{Probability density function of the slope of the radial ionised gas metallicity profile for galaxies at $z\approx 0.3$ for the ``well-resolved'' MAGPI-like samples of the cosmological hydrodynamical simulations {\sc EAGLE}, {\sc Magneticum}, {\sc HORIZON-AGN} and {\sc Illustris-TNG100}, as labelled in each panel. Here, the slope of the radial metallicity profile was measured at $r<r_{\rm e}$, with $r_{\rm e}$ being the half-light radius in the r-band.
We show separately the expected distribution in high and low-density environments, defined in the same way as in Fig.~\ref{fig:lambdaRevol}. The range of each histogram represents the Poisson error.}\label{fig:metalgradsenv}
\end{figure}

\begin{figure}
\includegraphics[trim=17mm 9.5mm 6mm 7mm, clip,width=0.245\textwidth]{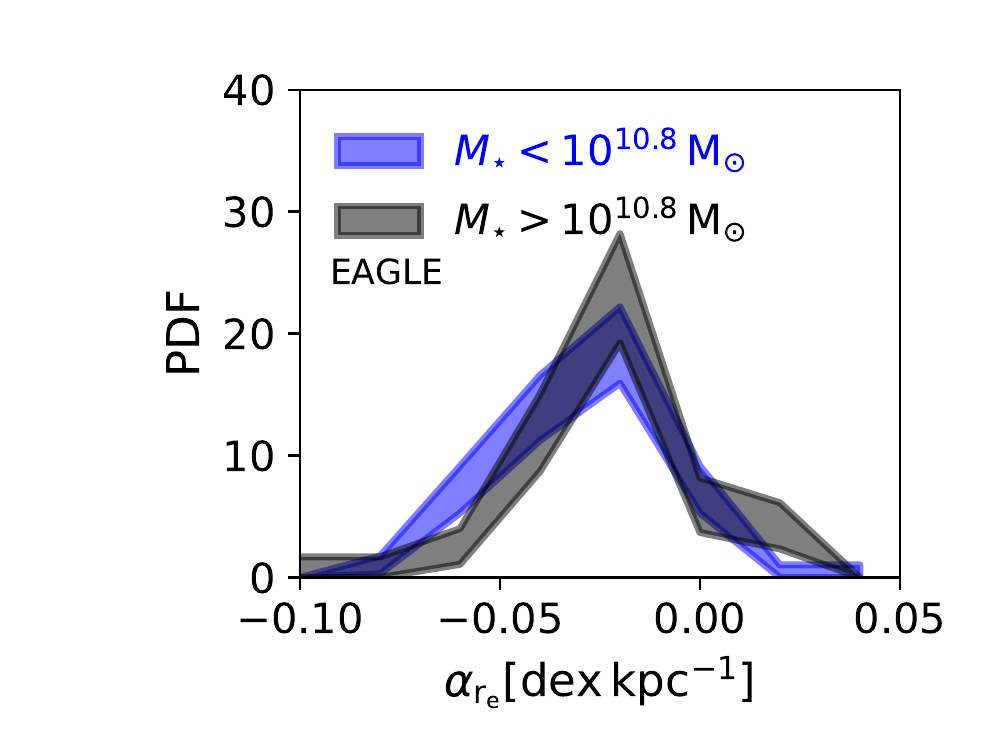}
\includegraphics[trim=22.5mm 9.5mm 6mm 7mm, clip,width=0.228\textwidth]{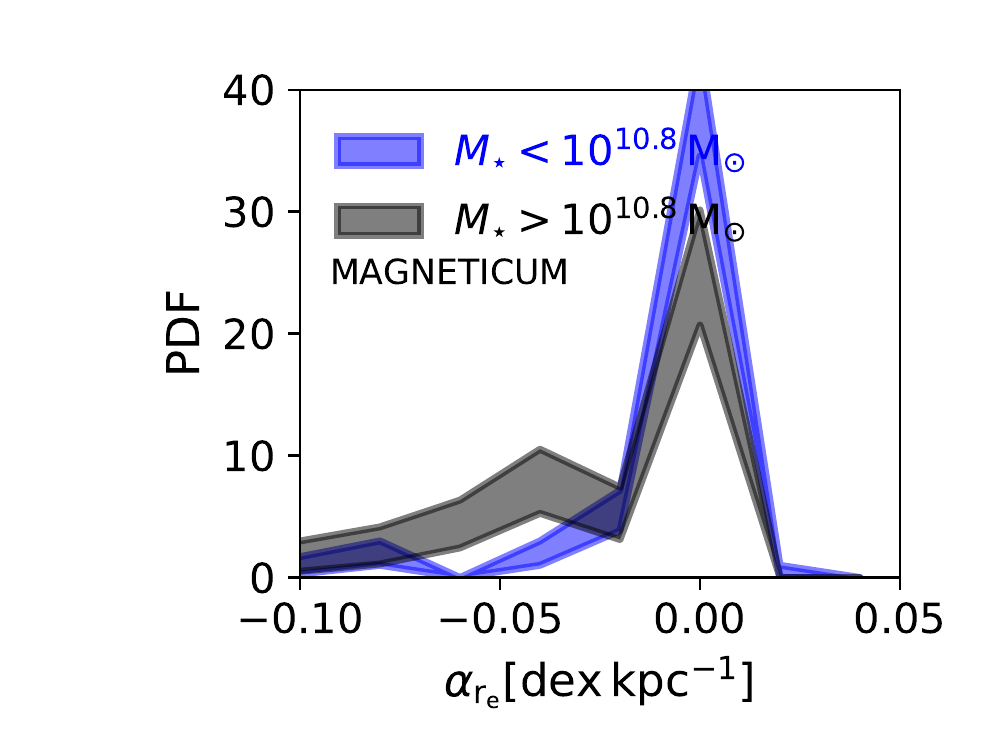}
\includegraphics[trim=17mm 4mm 6mm 7mm, clip,width=0.245\textwidth]{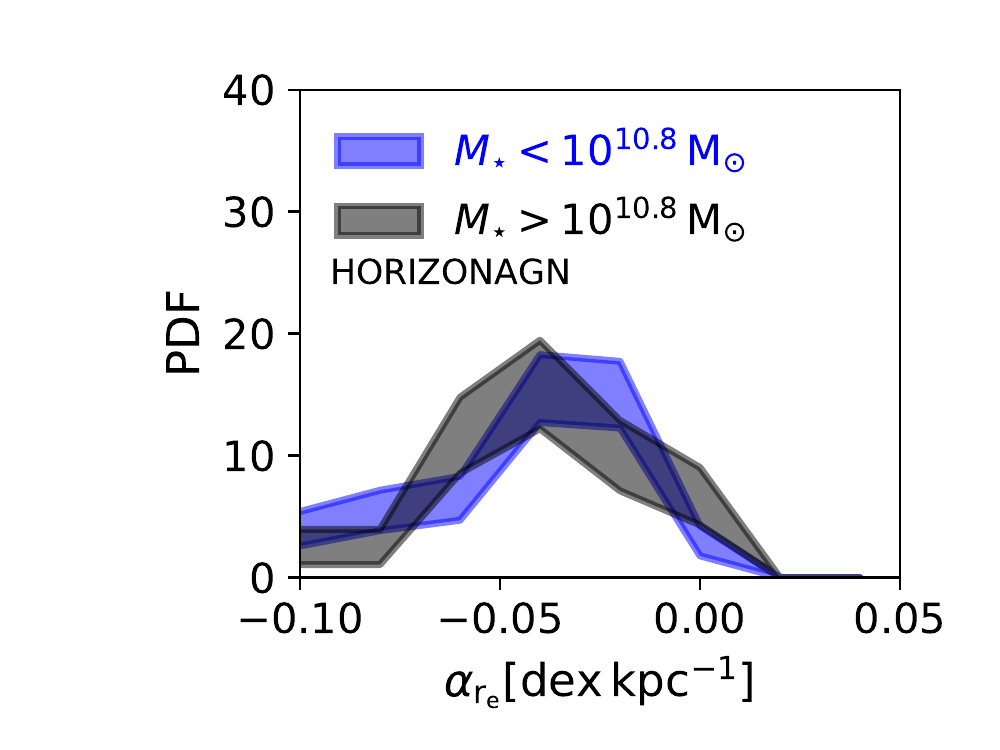}
\includegraphics[trim=22.5mm 4mm 6mm 7mm, clip,width=0.228\textwidth]{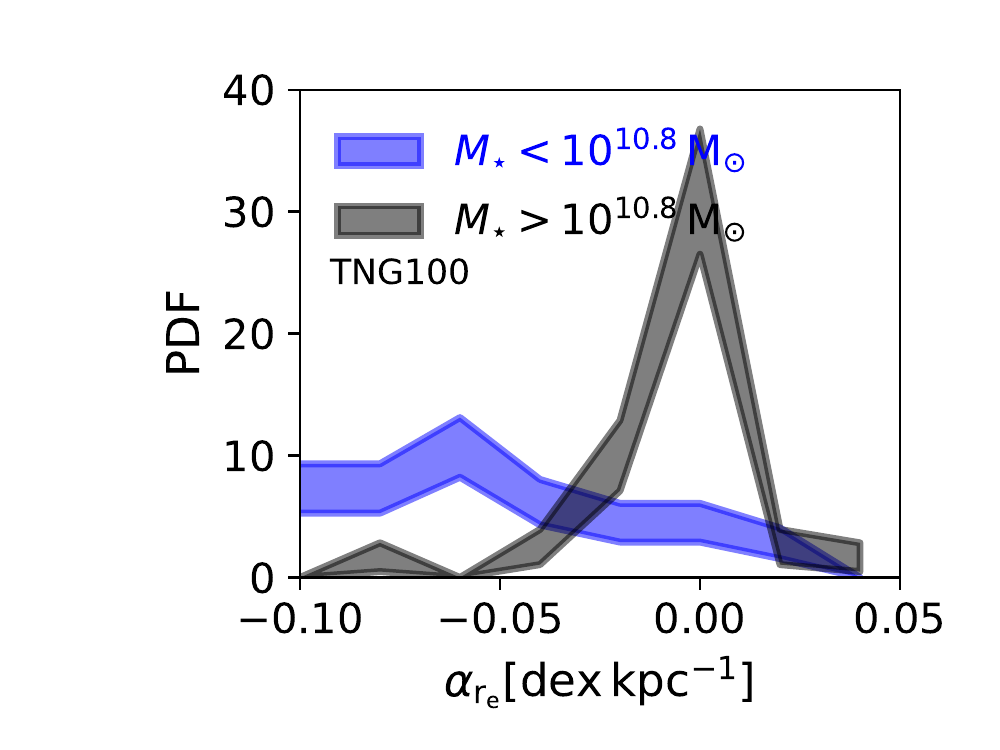}
\caption{As in Fig.~\ref{fig:metalgradsenv} but for two bins of stellar mass, as labelled.}\label{fig:metalgradsstellarmass}
\end{figure}
\begin{figure}
\includegraphics[trim=17mm 9.5mm 6mm 7mm, clip,width=0.245\textwidth]{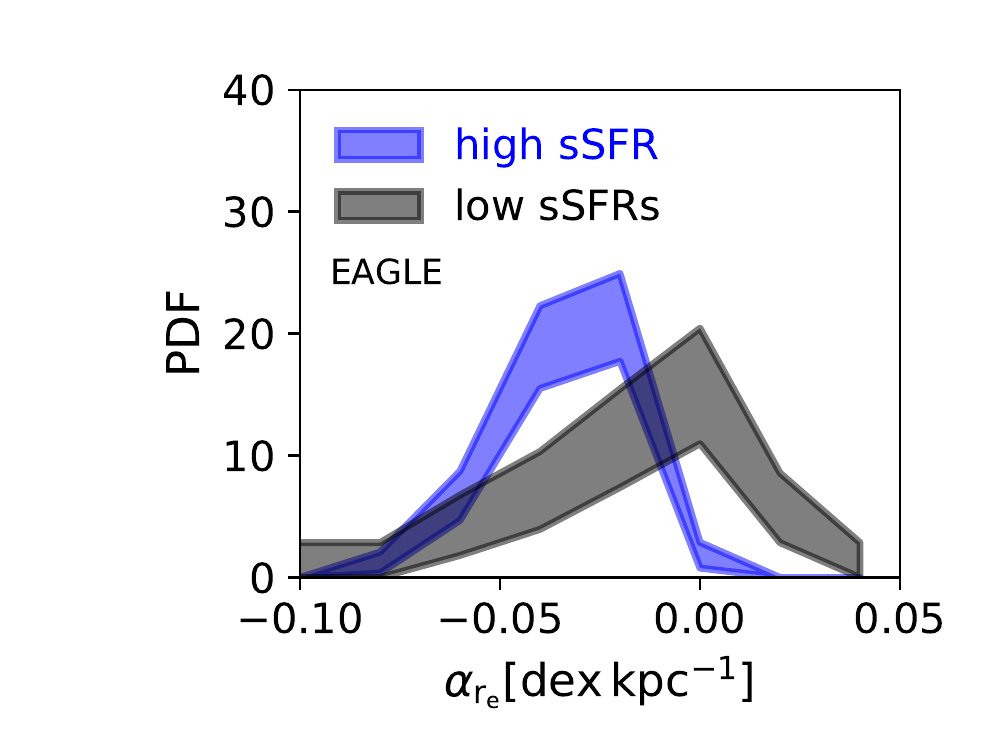}
\includegraphics[trim=22.5mm 9.5mm 6mm 7mm, clip,width=0.228\textwidth]{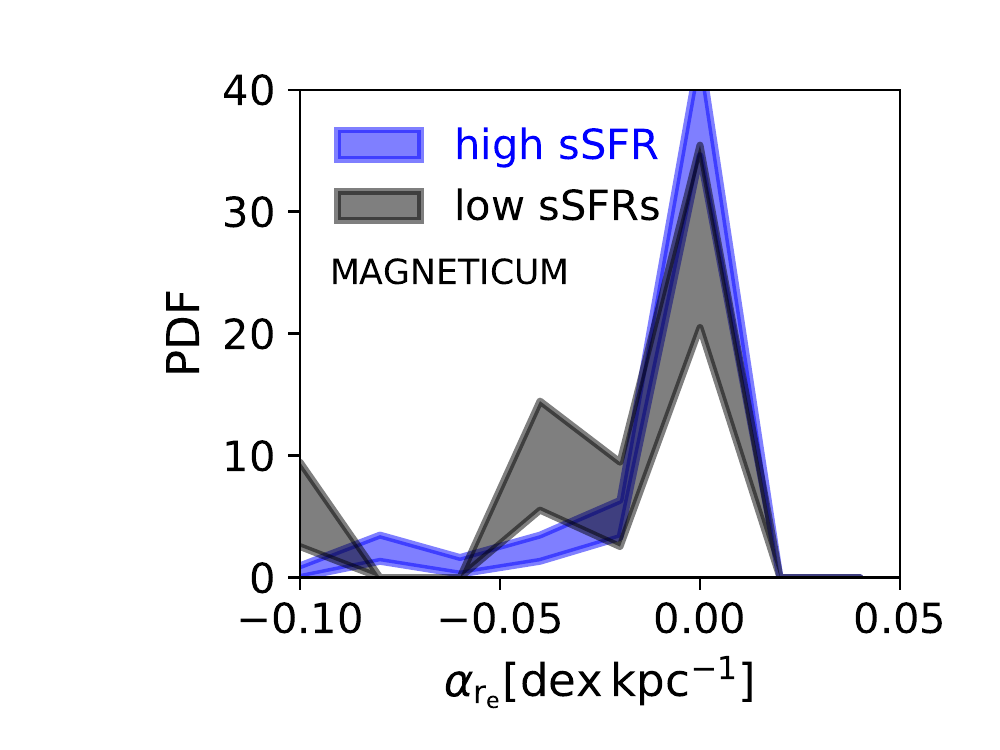}
\includegraphics[trim=17mm 4mm 6mm 7mm, clip,width=0.245\textwidth]{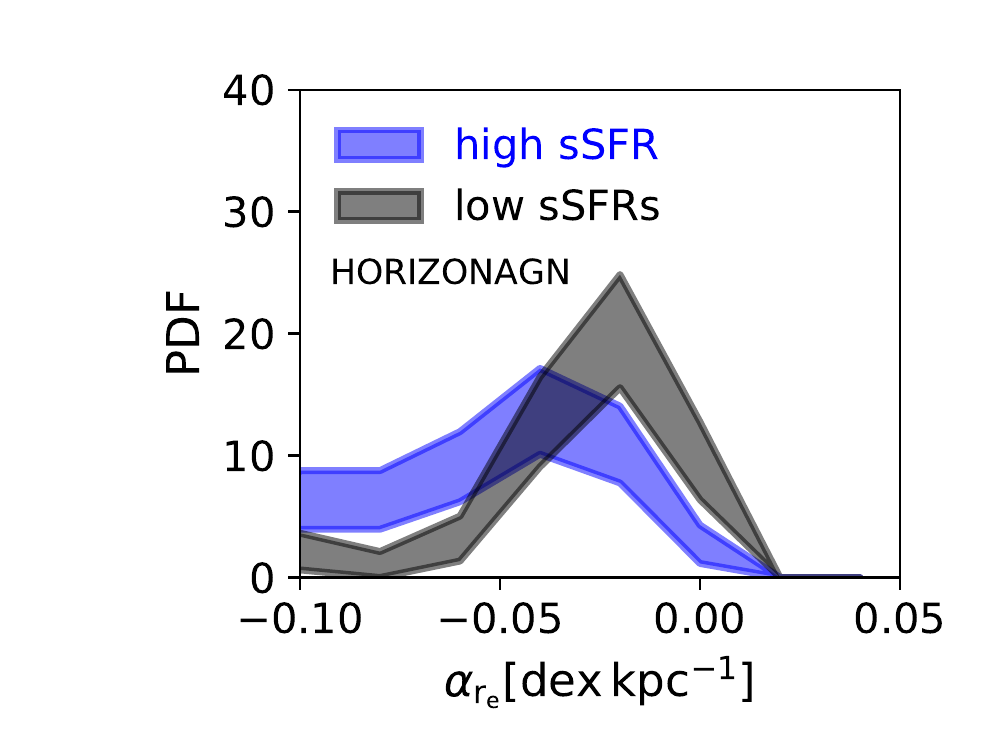}
\includegraphics[trim=22.5mm 4mm 6mm 7mm, clip,width=0.228\textwidth]{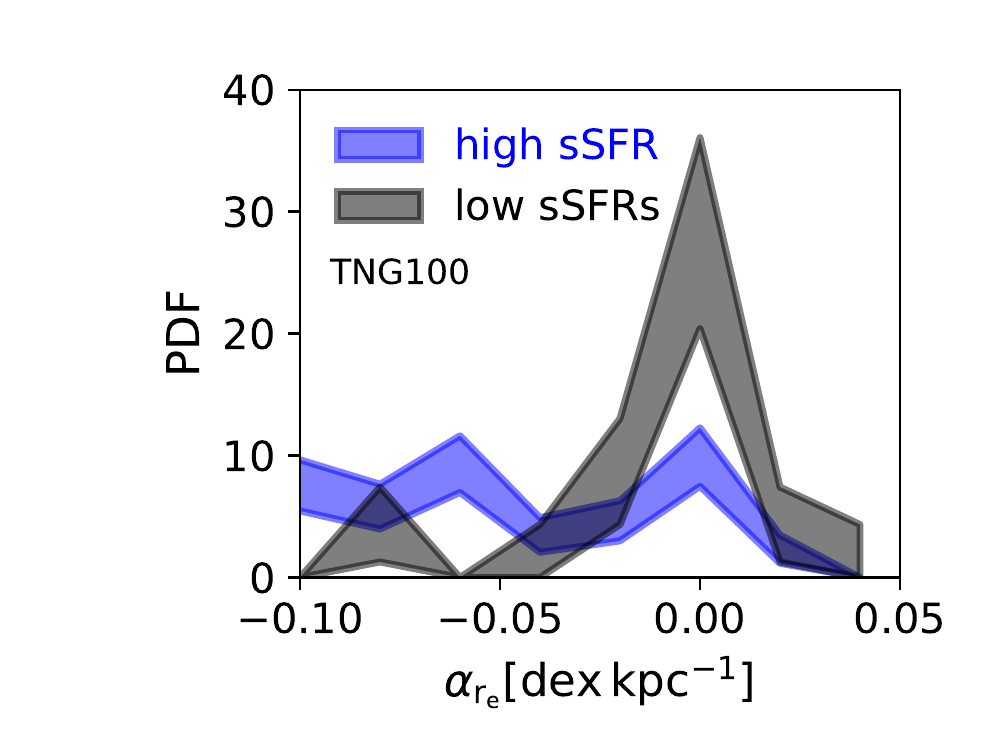}
\caption{As in Fig.~\ref{fig:metalgradsenv} but for galaxies in the lowest and highest $33^{\rm th}$ percentiles of the specific star formation rate distribution, as labelled.}\label{fig:metalgradsssfr}
\end{figure}

We explore the metallicity gradients in detail for the redshift of MAGPI, $z\sim 0.3$, using the ``well-resolved MAGPI-like'' samples from the cosmological hydrodynamical simulations {\sc EAGLE}, {\sc Magneticum}, {\sc HORIZON-AGN} (shown in Figs.~\ref{fig:lambdaRevol} and \ref{fig:lambdaRAtMAGPI}) and {\sc Illustris-TNG100}. Here, we do not show predictions from the TK15 simulation, as the statistics are not sufficient to build a full ``well-resolved MAGPI-like'' sample.
%Section~\ref{sec:theorycase} presents a description of {\sc Illustris-TNG100}.
%Specifications of the former three were provided in Section~\ref{sec:goal1}. {\sc Illustris-TNG100} includes subgrid physics modules for the same processes as the other simulations, but again that differ in the way modelling is done.  
%The cosmological volume of the simulation is $(100\,\rm Mpc)^3$ and using an unstructured mesh to solve the hydrodynamics equations. 
We study the diversity of radial metallicity profiles of the ionised gas in these four simulations and the predicted dependence on environment, stellar mass and specific star formation rate. %We do this by using the slope of the radial metallicity profile of the star-forming gas cells or particles, $\alpha_{\rm r_{50}}$ (see Section~\ref{section:datasims} for details on how this is computed).

Fig.~\ref{fig:metalgradsenv} shows the probability density function of $\alpha_{\rm r_{\rm e}}$ for the four simulations and in two density environments, which are defined in the same way as above and Fig.~\ref{fig:lambdaRevol}. Interestingly the simulations differ significantly in their predictions, with {\sc Illustris-TNG100} predicting a much wider distribution of $\alpha_{\rm r_{\rm e}}$ compared to the other simulations, also predicting a large fraction of positive $\alpha_{\rm r_{\rm e}}$; {\sc Magneticum}, on the other hand, predicts the narrowest distribution. 
In addition, {\sc HORIZON-AGN} and {\sc EAGLE} predict a clear environmental trend, while {\sc Magneticum} and {\sc Illustris-TNG100} predict a very weak or no environmental dependence. 

We study the effect of environment in these four simulations at $0\le z\le 1$ (not shown here) and find that each simulation predicts a different evolution of the environment dependence: {\sc EAGLE} predicts the environmental dependence to become weaker from $z=1$ to $z=0$, by which time environmental differences are very weak (see also \citealt{Tissera19}); in {\sc HORIZON-AGN}, there are very weak or no environmental differences in $\alpha_{\rm r_{\rm e}}$ at $z=1$ but those become more pronounced to $z=0$ (opposite to the {\sc EAGLE} prediction); while {\sc Magneticum} and {\sc Illustris-TNG100} predict environment to have a weak or no effect on $\alpha_{\rm r_{\rm e}}$ over the whole redshift range. 
The well defined environment metrics of MAGPI (as improved from GAMA) and the number of expected galaxies in these different environments will allow the survey to place stringent constraints on these predictions at $z\approx 0.3$, and combined studies of MAGPI together with local and more distant Universe surveys will allow to probe the evolutionary trends discussed here.

The presence or absence of an environmental trend of $\alpha_{\rm r_{\rm e}}$ in the well-resolved MAGPI sample of the four simulations is intimately linked with how these simulations predict $\alpha_{\rm r_{\rm e}}$ to vary with stellar mass and specific star formation rate. Figs.~\ref{fig:metalgradsstellarmass} and \ref{fig:metalgradsssfr} show the distribution of $\alpha_{\rm r_{\rm e}}$ for galaxies above and below a stellar mass of $10^{10.8}\rm \, M_{\odot}$ (the threshold used to define the primary MAGPI-like sample) and in the bottom and top $33^{\rm rd}$ percentile of specific star formation rate, respectively. We again see striking differences between the predictions: {\sc EAGLE} and  {\sc HORIZON-AGN} predict little dependence of $\alpha_{\rm r_{\rm e}}$ on stellar mass, while {\sc Magneticum} and {\sc Illustris-TNG100} predict a relatively weak and very strong stellar mass dependence, respectively. {\sc EAGLE} and {\sc HORIZON-AGN} predict a strong dependence of $\alpha_{\rm r_{\rm e}}$ on the specific star formation rate, while {\sc Magneticum} predicts a much weaker or no dependence on specific star formation rate. {\sc Illustris-TNG100} predicts the high specific star formation rates to have a significant tail towards very negative $\alpha_{\rm r_{\rm e}}$.
%The results for {\sc Illustris-TNG100} agree with those in \citet{Hemler20} for the higher resolution {\sc Illustris-TNG50} run of the suite, showing that this behavior is not resolution-driven and instead a result of the physics implementation in {\sc Illustris-TNG}. 

The dependence of $\alpha_{\rm r_{\rm e}}$ on specific star formation rate in {\sc EAGLE} is a direct consequence of the effect gas accretion has on $\alpha_{\rm r_{\rm e}}$ as described in \citet{Collacchioni20}, and hence the differences here suggest that gas accretion may be having a lesser or at least different role in modifying metal mixing within galaxies in some of the other simulations included here (e.g. {\sc Illustris-TNG100} and {\sc Magneticum}). It is difficult to pinpoint the exact cause for these differences, but we speculate that the fact that outflows behave very differently in these simulations is a likely culprit. Stellar feedback in {\sc EAGLE} is very effective at removing gas from lower mass galaxies and halos \citep{Davies20,Mitchell20} and even significantly decreasing further gas accretion onto halos \citep{Wright20}, though at higher masses this role is overtaken by AGN feedback. In contrast, outflows in {\sc Illustris-TNG100}, for example, generally do not lead to gas escaping from halos, leading to quick reincorporation of the outflowing gas \citep{Nelson19b,Mitchell20}. 
%Yingjie Peng:
In addition to gas accretion and outflows driven by feedback, galaxies with higher velocity dispersions in gas are observed to have shallower or more positive metallicity gradients \citep{Queyrel12}, as the velocity dispersion helps to radially mix the gas and metals \citep{Krumholz2018,Hemler20, Sharda21}.

The fact that there are such discrepant predictions among cosmological hydrodynamical simulations presents a great opportunity for MAGPI to place powerful constraints and start to identify areas of tension with the simulations that can hopefully lead to further development in galaxy formation theory. Furthermore, since some of the simulations are calibrated to reproduce specific ``global'' observables (e.g., galaxy size, black hole mass, stellar mass distribution, etc.), spatially-resolved properties offer a greater opportunity to really break the degeneracy between different physical models.

We highlight that the expected well-resolved structures of galaxies in the MAGPI primary sample will allow much more detailed studies than the general trends explored here with cosmological hydrodynamical simulations. We will look for the evolution (by comparison with local surveys) of possible links between morphological features such as bars, spiral arms and disturbances indicative of galaxy mergers and the 2-dimensional metallicity information of massive galaxies to understand the role of dynamical features on the metal mixing of gas in galaxies \citep{Kreckel19, Zurita21}.

\section{Summary}\label{section:summary}

Galaxies have undergone significant dynamical and morphological evolution over the last 8 billion years of cosmic time ($0 < z < 1$). During this epoch, the overall star formation activity of galaxies shows a steep decline in both volume density, and at fixed stellar mass \citep{MadauDickinson14}.

The \underline{M}iddle \underline{A}ges \underline{G}alaxy \underline{P}roperties with \underline{I}FS (MAGPI) survey is designed to efficiently probe galaxy transformation and the role of nature vs. nurture, by combining spatially-resolved IFS data with robust environmental metrics at intermediate redshift ($z \sim 0.3$). MAGPI is a VLT/MUSE Large Program to obtain resolved observations of gas and stars at $z=0.25-0.35$ in 60 galaxies in a representative range of environments (halo mass) and up to 100 of their neighbouring satellites. MAGPI will also obtain unresolved spectra for a further $\sim50$ satellites. In addition to observations, the MAGPI survey tightly integrates theoretical models and simulations, including a detailed plan for the production of mock observations.

MAGPI fills a so-far-unexplored region of parameter space (Fig. \ref{fig:spatialres}): in terms of mapping the properties of stars, it pierces farther than any of the local surveys (e.g. MaNGA, SAMI, CALIFA, ATLAS$^{\rm 3D}$, etc.); in terms of mapping gas properties, like K-CLASH it bridges the gap between said surveys and the high-redshift gas-only IFS samples (e.g. KMOS3D, IMAGES, MASSIV, SINS, etc.).

The primary goal of the MAGPI survey is to reveal and understand the physical processes responsible for the rapid transformation of galaxies at intermediate redshift by:
\begin{itemize}
\item detecting the impact of environment (\S~\ref{sec:goal1});
\item understanding the role of gas accretion and merging (\S~\ref{sec:goal2});
\item determining energy sources and feedback activity (\S~\ref{sec:goal3});
\item tracing the metal mixing history of galaxies (\S~\ref{sec:goal4}); and
\item producing a comparison-ready theoretical dataset (\S~\ref{sec:theorycase}).
\end{itemize}

The MAGPI survey design, strategy and data handling are chosen to address the above science goals (see \S~\ref{sec:data}). The observational campaign is ongoing, with four MUSE fields having so far been at least partly observed, and two publicly available archive fields. \S~\ref{section:results} showcases some of the early observational and theoretical results of the survey to date, including stellar population measurements and maps (Fig. \ref{fig:fieldwithinsets}), gas-phase metallicity and kinematic maps (Fig. \ref{fig:gasandstars}), stellar kinematic maps (both observed, Figs. \ref{fig:fieldwithinsets} and \ref{fig:gasandstars}; and simulated, Fig. \ref{fig:examplemaps}), theoretical predictions for the impact of environment, stellar mass and star formation on metallicity gradients (Figs. \ref{fig:metalgradsenv}, \ref{fig:metalgradsstellarmass} and \ref{fig:metalgradsssfr}) and stellar spin (Figs. \ref{fig:lambdaRevol} and \ref{fig:lambdaRAtMAGPI}) at $z\sim0.3$ and over cosmic time.

The MAGPI team is committed to a collaborative approach to achieve the survey science goals stated above. This entails regular data releases to maximise community involvement. See the MAGPI Survey webpage \href{https://magpisurvey.org}{\url{https://magpisurvey.org}} for further and up-to-date information.

\begin{acknowledgements}
%ESO
Based on observations collected at the European Organisation for Astronomical Research in the Southern Hemisphere under ESO program 1104.B-0536. We wish to thank the ESO staff, and in particular the staff at Paranal Observatory, for carrying out the MAGPI observations. 
%ASTRO3D
Part of this research was conducted by the Australian Research Council Centre of Excellence for All Sky Astrophysics in 3 Dimensions (ASTRO 3D), through project number CE170100013. 
%GAMA
MAGPI targets were selected from GAMA. GAMA is a joint European-Australasian project based around a spectroscopic campaign using the Anglo-Australian Telescope. GAMA is funded by the STFC (UK), the ARC (Australia), the AAO, and the participating institutions. GAMA photometry is based on observations made with ESO Telescopes at the La Silla Paranal Observatory under programme ID 179.A-2004, ID 177.A-3016. 
%Magneticum
The {\it Magneticum} Pathfinder simulations were performed at the Leibniz-Rechenzentrum with CPU time assigned to the Project {\it pr86re} and supported by the Deutsche Forschungsgemeinschaft (DFG, German Research Foundation) under Germany's Excellence Strategy - EXC-2094 - 390783311.
%Phil
The simulation presented in \citet{Taylor17} was run on the University of Hertfordshire's high-performance computing facility.
%Chiaki
CK acknowledges funding from the UK Science and Technology Facility Council (STFC) through grant ST/M000958/1 \& ST/ R000905/1, and the Stromlo Distinguished Visitorship at the ANU. CK's work used the DiRAC Data Centric system at Durham University, operated by the Institute for Computational Cosmology on behalf of the STFC DiRAC High Performance Computing (HPC) Facility (\url{https://dirac.ac.uk}). This equipment was funded by a BIS National E- infrastructure capital grant ST/K00042X/1, STFC capital grant ST/ K00087X/1, DiRAC Operations grant ST/K003267/1 and Durham University. DiRAC is part of the National E-Infrastructure.
%Anna
AFM acknowledges support of the Postdoctoral Junior Leader Fellowship Programme from 'La Caixa' Banking Foundation (LCF/BQ/LI18/11630007).
%Evelyn
EJJ acknowledges support from FONDECYT Postdoctoral Fellowship Project No.\,3180557 and FONDECYT Iniciaci\'on 2020 Project No.\,11200263.
%Florencia
FC acknowledges CONICET, Argentina, and the Australian Endeavour Scholarships and Fellowships for their supporting fellowships.
%Francesco
FDE acknowledges funding through the H2020 ERC Consolidator Grant 683184.
%Jesse
JvdS acknowledges support of an Australian Research Council Discovery Early Career Research Award (project number DE200100461) funded by the Australian Government.
%Piyush
PS is supported by the Australian Government Research Training Program (RTP) Scholarship.
%Richard
RMcD is the recipient of an Australian Research Council Future Fellowship (project number FT150100333).
%Sabine
SB acknowledges support from the Australian Research Council under Discovery Project Discovery 180103740.
%Sukyoung
SKY acknowledges support from the Korean National Research Foundation (NRF-2020R1A2C3003769).
%Tania
TMB is supported by an Australian Government Research Training Program Scholarship.
%Yingjie
YP acknowledges the National Key R\&D Program of China, Grant 2016YFA0400702 and National Science Foundation of China (NSFC) Grant No. 11773001, 11721303, 11991052.
%CMasher for good-lookin' colour scales.
This work makes use of colour scales chosen from \citet{CMASHER2020}. 
\end{acknowledgements}

\begin{appendix}

\section{MAGPI target list}

This section presents information about the MAGPI Survey fields in Table \ref{tab:targetlist2} and postage-stamp KIDS images in Figs. \ref{fig:fieldsgallery}. Synthetic images of the archive fields Abell 370 and Abell 2477 are shown in Fig. \ref{fig:archivefields}.

\onecolumn
\begin{center}
\begin{landscape}
\begin{longtable}{lcccccrrcc}
\caption{List of MAGPI fields and primary object properties. Column (1): Field name. Column (2): unique MAGPI field ID. Column (3): primary object GAMA CATAID. Column(4): primary object redshift, derived from GAMA. Column (5) and Column (6): right ascension and declination of primary object. Note that this does not necessarily correspond to the field centre. Column(7): $g-i$ colour from KiDS. Column(8): half-light size derived following \citet{Kelvin12}. Column(9): galaxy stellar mass from \citet{Taylor11}. Column (10): dark matter halo mass, taken from the G3C catalogues described in \citet{Robotham11}.}\label{tab:targetlist2} \\
\hline Field name & FieldID & GAMA CATAID & $z$ & R.A. & Decl.  & ${g-i}$ & $R_\mathrm{e}$ & $\log (M_\star/M_{\odot})$ & $\log (M_\mathrm{halo}/M_{\odot})$\\
  & & (primary) & (primary) & (J2000) & (J2000) & & (kpc) & & \\
 (1) & (2) & (3) & (4) & (5)  & (6) & (7) & (8) & (9) & (10) \\ \hline 
\endfirsthead
\hline \multicolumn{10}{l}{{Continued on next page}} \\
\endfoot
\multicolumn{10}{c}%
{{\bfseries \tablename\ \thetable{} -- continued from previous page}} \\
\hline Field name & FieldID & GAMA CATAID & $z$ & R.A. & Decl. & ${g-i}$ & $R_\mathrm{e}$ & $\log (M_\star/M_{\odot})$ & $\log (M_\mathrm{halo}/M_{\odot})$\\
  & & (primary) & (primary) & (J2000) & (J2000)  & & (kpc) & &  \\
 (1) & (2) & (3) & (4) & (5)  & (6) & (7) & (8) & (9) & (10) \\ \hline 
\endhead
\hline \hline
\endlastfoot
J113850  & 1201 &   176902 & 0.3469 & 174.7112 &  -1.9282 &  1.10 & 12.92 & 11.57 & 14.56\\
J114121  & 1202 &   184180 & 0.2918 & 175.3388 &  -1.5823 &  0.72 & 11.52 & 10.90 & 14.78\\
J114123  & 1203 &     7043 & 0.3097 & 175.3473 &   0.6337 &  1.09 & 13.97 & 11.76 & 14.58\\
J114238  & 1204 &    39176 & 0.3158 & 175.6612 &  -0.7943 &  0.89 & 10.54 & 11.07 & 12.74\\
J115219  & 1205 &    39777 & 0.2914 & 178.0798 &  -0.8268 &  0.88 &  8.29 & 11.00 & 12.56\\
J120038  & 1206 &   185407 & 0.3265 & 180.1619 &  -1.4520 &  1.09 & 22.54 & 11.47 & 13.81\\
J120759  & 1207 &   172252 & 0.3206 & 181.9999 &  -2.4836 &  0.77 &  7.86 & 10.96 & 13.61\\
J121953  & 1208 &   172929 & 0.3005 & 184.9716 &  -2.4810 &  1.03 &  7.57 & 11.37 & 15.11\\
J122223  & 1209 &   145672 & 0.2959 & 185.5985 &  -1.3800 &  0.96 &  8.23 & 10.96 & 12.46\\
J140913  & 1501 &   237785 & 0.3095 & 212.3053 &   1.7832 &  1.01 & 23.24 & 11.51 & 13.16\\
J141031  & 1502 &   260943 & 0.2967 & 212.6327 &   2.5610 &  1.06 &  8.65 & 11.20 & 13.83\\
J141428  & 1503 &    62746 & 0.2877 & 213.6190 &  -0.4152 &  0.95 &  6.19 & 10.92 & 13.13\\
J141429  & 1504 &   319143 & 0.3060 & 213.6243 &   1.9707 &  0.78 &  8.93 & 11.13 & 13.58\\
J141837  & 1505 &   507859 & 0.3175 & 214.6580 &  -1.7180 &  0.99 &  6.44 & 10.90 & 11.80\\
J142109  & 1506 &   238775 & 0.2966 & 215.2885 &   1.6235 &  1.03 &  6.43 & 10.98 & 12.28\\
J142228  & 1507 &   618422 & 0.3146 & 215.6206 &   0.4075 &  0.95 & 11.37 & 11.07 & 13.50\\
J142332  & 1508 &   362622 & 0.3159 & 215.8836 &   2.7140 &  1.01 & 11.76 & 10.86 & 12.47\\
J142333  & 1509 &   362613 & 0.2829 & 215.8913 &   2.6404 &  1.04 & 26.47 & 11.76 & 14.98\\
J142506  & 1510 &   250872 & 0.2865 & 216.2763 &   2.1173 &  0.92 &  5.40 & 10.93 & 14.35\\
J142617  & 1511 &   320018 & 0.2939 & 216.5714 &   1.7308 &  1.27 & 12.33 & 11.29 & 12.74\\
J142620  & 1512 &   485594 & 0.3213 & 216.5863 &  -1.7174 &  1.02 &  8.52 & 11.10 & 13.50\\
J142858  & 1513 &   320174 & 0.3149 & 217.2457 &   1.7331 &  0.98 & 20.99 & 11.28 & 14.60\\
J142859  & 1514 &   362922 & 0.3025 & 217.2460 &   2.6123 &  0.95 &  7.78 & 10.99 & 14.47\\
J143127  & 1515 &   297976 & 0.2878 & 217.8630 &   1.3771 &  0.99 &  8.14 & 11.14 & 12.56\\
J143154  & 1516 &   298034 & 0.3353 & 217.9750 &   1.3036 &  1.10 & 10.21 & 11.15 & 13.07\\
J143215  & 1517 &   508727 & 0.2867 & 218.0628 &  -1.6101 &  1.08 &  8.89 & 11.26 & 13.15\\
J143234  & 1518 &    64101 & 0.3338 & 218.1451 &  -0.2171 &  0.99 & 12.30 & 11.60 & 12.28\\
J143242  & 1519 &   512241 & 0.3340 & 218.1765 &  -1.1139 &  1.08 & 14.20 & 11.44 & 13.89\\
J143422  & 1520 &   569278 & 0.3385 & 218.5933 &  -0.4336 &  0.83 &  5.34 & 10.96 & 13.56\\
J143512  & 1521 &   492903 & 0.2889 & 218.8001 &  -1.3569 &  1.00 & 14.32 & 11.54 & 14.44\\
J143616  & 1522 &    16528 & 0.2936 & 219.0677 &   0.8004 &  0.90 &  4.85 & 10.88 & 12.86\\
J143809  & 1523 &   512647 & 0.2807 & 219.5409 &  -1.0993 &  1.04 &  9.24 & 11.31 & 13.55\\
J143836  & 1524 &   512697 & 0.3302 & 219.6541 &  -1.0501 &  0.97 & 13.03 & 11.44 & 14.12\\
J143840  & 1525 &   619409 & 0.3181 & 219.6672 &   0.3336 &  0.99 &  8.99 & 11.23 & 13.67\\
J143918  & 1526 &   343323 & 0.3025 & 219.8265 &   2.2530 &  0.86 & 13.38 & 11.02 & 12.89\\
J144010  & 1527 &    49219 & 0.2904 & 220.0441 &  -0.6566 &  1.13 &  6.34 & 10.94 & 13.78\\
J144055  & 1528 &   486569 & 0.3225 & 220.2323 &  -1.6481 &  0.94 & 10.55 & 11.00 & 14.21\\
J144128  & 1529 &   594500 & 0.3455 & 220.3669 &  -0.0995 &  1.01 &  6.06 & 11.07 & 13.81\\
J144834  & 1530 &   367422 & 0.3100 & 222.1438 &   2.9413 &  1.07 &  6.27 & 11.26 & 14.01\\
J145136  & 1531 &   487117 & 0.3466 & 222.9035 &  -1.6821 &  1.03 & 21.06 & 11.48 & 14.25\\
J145150  & 1532 &    79765 & 0.3189 & 222.9613 &   0.0997 &  1.00 &  7.61 & 11.19 & 13.55\\
J145152  & 1533 &   595015 & 0.3153 & 222.9690 &  -0.1419 &  0.98 &  6.83 & 11.00 & 13.23\\
J145221  & 1534 &   546078 & 0.3130 & 223.0899 &  -0.9722 &  1.02 & 21.95 & 11.44 & 14.79\\
J145231  & 1535 &   595037 & 0.3142 & 223.1313 &  -0.0718 &  1.03 &  9.69 & 11.22 & 11.83\\
J223757  & 2301 &  5103706 & 0.2984 & 339.4911 & -31.8550 &  1.08 & 11.00 & 11.37 & 14.72\\
J224045  & 2302 &  5209008 & 0.2933 & 340.1908 & -34.7149 &  1.08 & 15.59 & 11.33 & 11.36\\
J224128  & 2303 &  5212548 & 0.3429 & 340.3697 & -34.0800 &  0.86 & 11.97 & 11.01 & 13.35\\
J224634  & 2304 &  5121184 & 0.2858 & 341.6425 & -31.3054 &  1.14 &  7.22 & 11.04 & 14.02\\
J225825  & 2305 &  5252712 & 0.3172 & 344.6067 & -34.6966 &  0.97 & 12.62 & 10.89 & 13.88\\
J230015  & 2306 &  5273998 & 0.3138 & 345.0657 & -34.4678 &  1.02 &  7.76 & 10.95 & 14.32\\
J230156  & 2307 &  5273865 & 0.3356 & 345.4847 & -34.4975 &  1.05 & 12.97 & 11.29 & 12.89\\
J230158  & 2308 &  5154123 & 0.3470 & 345.4921 & -32.7987 &  0.96 &  8.88 & 11.16 & 14.53\\
J230506  & 2309 &  5286176 & 0.3260 & 346.2757 & -30.1207 &  0.95 &  7.74 & 10.99 & 13.68\\
J231312  & 2310 &  5316104 & 0.2839 & 348.3031 & -34.0145 &  0.93 & 13.63 & 10.98 & 12.63\\
J231349  & 2311 &  5320902 & 0.3323 & 348.4545 & -32.9826 &  1.05 &  5.77 & 10.99 & 13.78\\
J231911  & 2312 &  5341265 & 0.3385 & 349.7997 & -33.3578 &  0.89 &  9.27 & 10.96 & 12.93\\
\end{longtable}
\end{landscape}

\end{center}

\twocolumn

\begin{figure*}
\includegraphics[width=16cm]{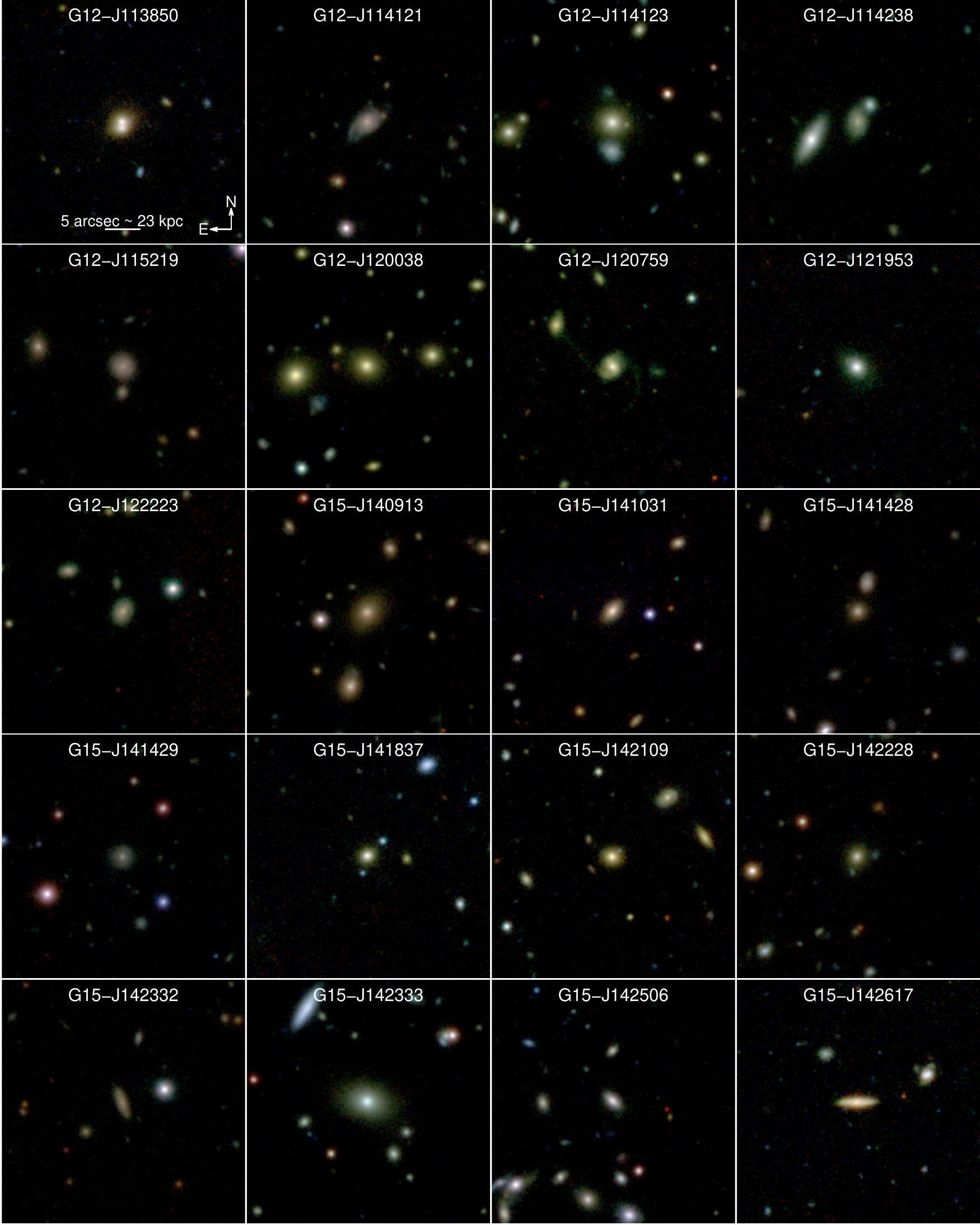}
\caption{Colour (${\rm R}=Z$, ${\rm G}=r$, ${\rm B}=g_{\rm mod}$) KiDS images for all MAGPI fields, as labelled. In all panels, and as labelled in the top left panel, North is up and East is to the left. A scale of 5 arcsec (corresponding to $\sim23$ kpc at $z\sim0.3$) is shown on the top left panel for reference. All square images are 1 arcmin to the side.}\label{fig:fieldsgallery}
\end{figure*}

\setcounter{figure}{11}
\begin{figure*}
\centering
\includegraphics[width=16cm]{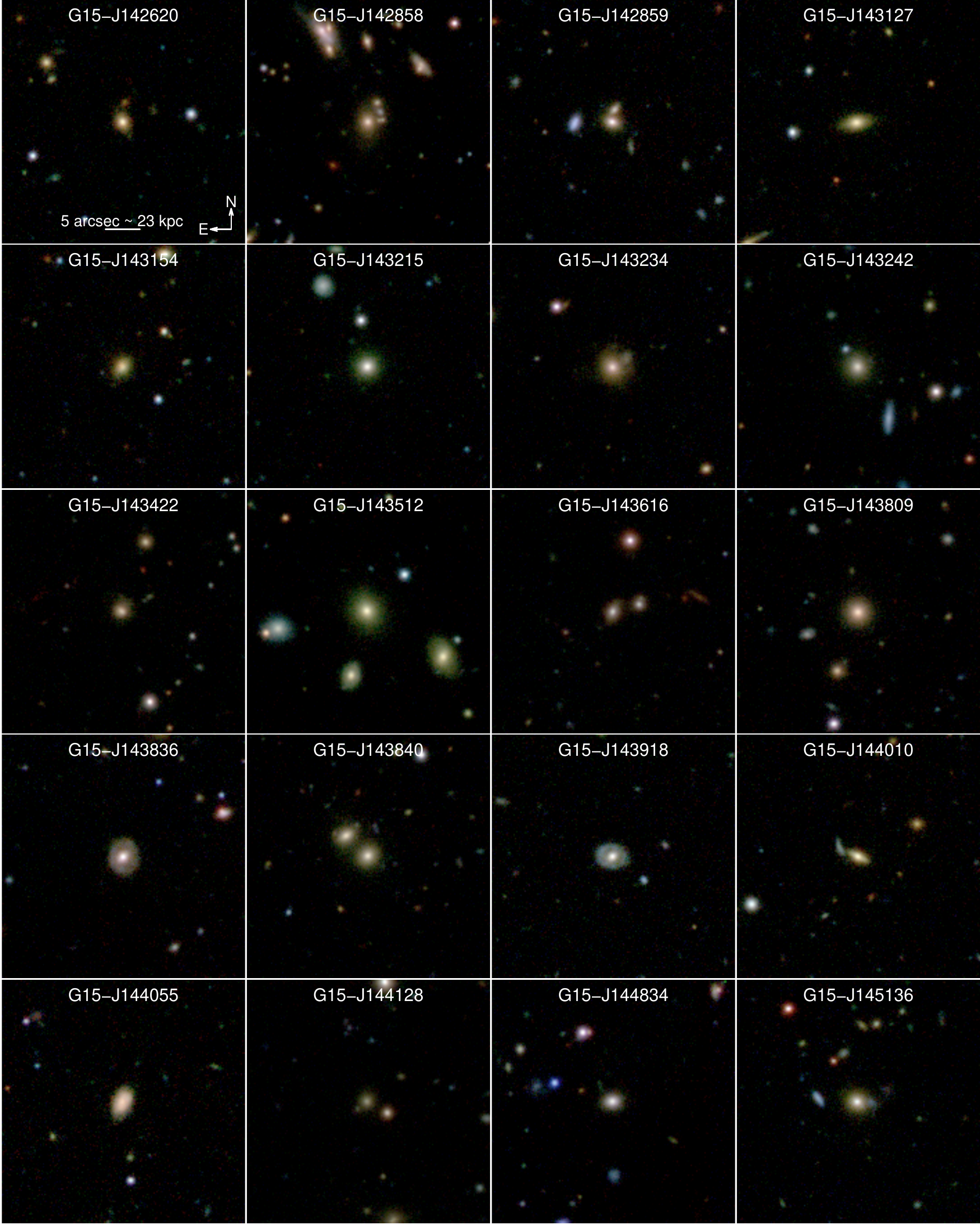}
\caption{continued.}
\end{figure*}

\setcounter{figure}{11}
\begin{figure*}
\includegraphics[width=16cm]{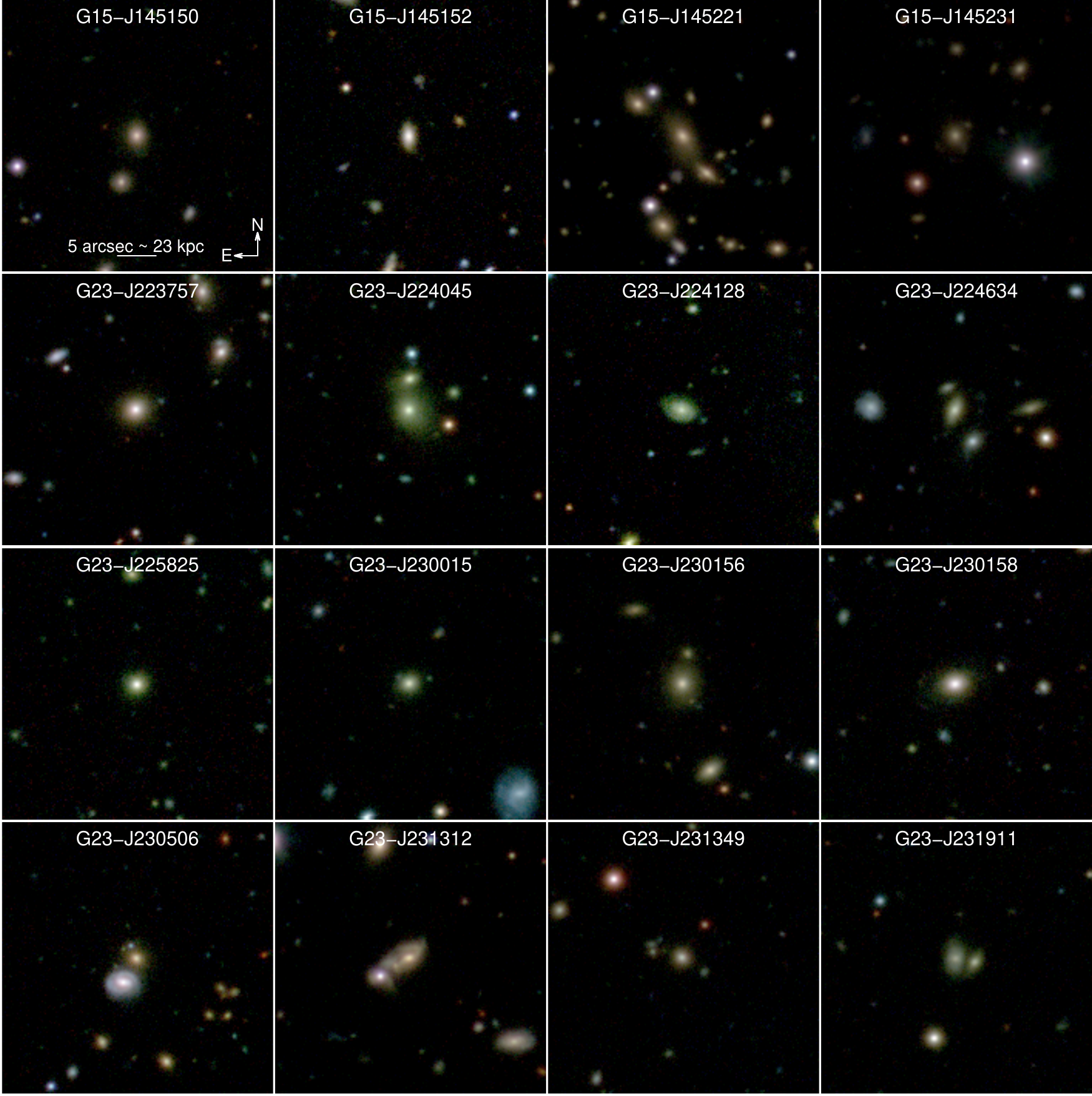}
\caption{continued.}
\end{figure*}

\begin{figure*}
\includegraphics[width=16cm]{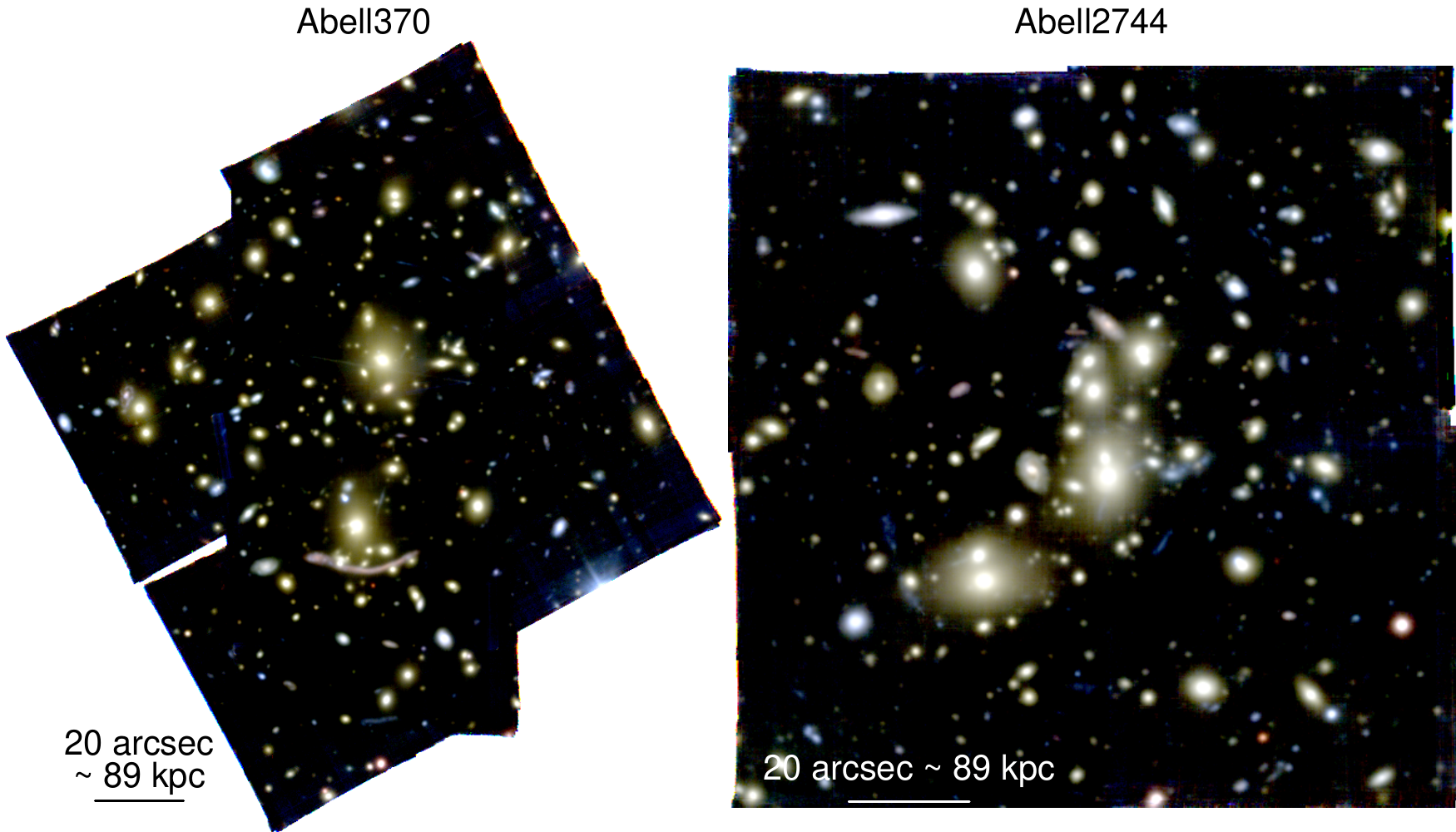}
\caption{Synthetic colour (${\rm R}=i$, ${\rm G}=r$, ${\rm B}=g_{\rm mod}$) image mosaics for the archive lensing cluster fields Abell 370 ($z=0.375$, left; Program ID 096.A-0710, PI: Bauer) and Abell 2477 ($z=0.308$, right; Program IDs 095.A-0181 and 096.A-0496, PI: Richard) based on available reduced data from the MUSE consortium (\citealt{Lagattuta19} and \citealt{Mahler18} for Abell 370 and Abell 2744, respectively). These archival data are used to probe the highest densities for the MAGPI survey. A scale of 20 arcsec ($\sim89$ kpc at $z\sim0.3$) is given on each panel for reference and in both cases North is pointing up and East to the left.}\label{fig:archivefields}
\end{figure*}

\section{Simulations description}\label{sec:simsdescription}
We provide a short summary of the simulations currently in our suite. We focus on processes that we consider key to MAGPI: gas cooling, interstellar medium modelling, star formation and feedback from stars and supermassive black holes, and defer the reader to the original papers for more details.

\noindent - {\sc EAGLE}\footnote{\url{http://icc.dur.ac.uk/Eagle/}}. Metal radiative cooling is included following \citet{Wiersma09}; the interstellar medium model of galaxies imposes a polytropic equation of state roughly when the gas cools down to $10^4$~K, to avoid the very short timescales typically associated with the dense gas. Stars form probabilistically from gas that is considered dense enough for its metallicity \citep{Schaye08}. Stellar feedback is modelled also probabilistically as energy injection to neighbouring particles, which heats them up to $10^{7.5}$~K. This temperature is hot enough to mitigate quick radiative losses and therefore aids the formation of galactic winds. AGN feedback is modelled in a similar way (and hence as a single heating mode), but the neighbouring gas particles are heated to a higher temperature of $10^{8.5}\rm \,K$. 

\noindent - {\sc Magneticum}\footnote{\url{http://www.magneticum.org}}. %Metal radiative cooling also follows \citet{Wiersma09} as in {\sc EAGLE} and {\sc Illustris-TNG100}; 
Metals and energy are released by stars of different mass by integrating the evolution of the stellar population \citep[see][for details]{Dolag17} (in a similar fashion to the other simulations summarised here).
%and account for mass-dependent lifetimes following \citet{Padovani93}), as well as metallicity-dependent SNe II \citep{Woosley95}, AGB stars \citep{vandenHoek97}, and SNe Ia \citep{Thielemann03} yields. 
%New formed stars follow a \citep{Chabrier03} initial mass function.
The interstellar medium is treated as a two-phase medium where clouds of cold gas form from cooling of hot gas and are embedded in the hot gas phase assuming pressure equilibrium whenever gas particles are above a given threshold density, $n_{\rm H}\approx 0.5\,\rm cm^{-3}$ \citep{Springel03}. This two-phase medium has a similar effect to the polytropic equation of state adopted in {\sc EAGLE}, in that very short timescales are avoided. Stars form probabilistically at densities in excess of the above density threshold. Stellar feedback is modelled as kinetic energy injection in the form of an isotropic wind that is decoupled from the hydrodynamic calculation for a period of time that is enough as to allow the particles to escape the local interstellar medium. AGN feedback also injects energy, but with two different efficiencies depending on whether black holes are above/below a given Eddington ratio (mimicking a two QSO/radio AGN feedback model).
%Contrary to all other simulations, 
Magneticum additionally follows thermal conduction, similar to \citet{Dolag04}, but with a choice of 1/20 of the classical Spitzer value \citep{Spitzer62, Arth14}.
%The choice of a suppression value significantly below 1/3 can be justified by comparison with full MHD simulations including an anisotropic treatment of thermal conduction (see discussion in \citealt{Arth14}).

\noindent - {\sc HORIZON-AGN}\footnote{\url{https://www.horizon-simulation.org}}. Metal radiative cooling is followed using the cooling tables of \citet{Sutherland93}. Gas can cool down to $10^4$~K, and stars form from gas above a density threshold of $n_{\rm H}\approx 0.1\,\rm cm^{-3}$. Feedback from stars is modelled as energy injection. AGN feedback is modelled as two modes: a QSO mode, which releases thermal energy in a similar way to how it is done in {\sc EAGLE} (though particles are heated to a lower temperature, $10^7$~K), and a jet mode, that deposits mass, energy and momentum to a small cylinder into the interstellar medium (which ultimately mimics a bipolar outflow). Whether a black hole is capable of QSO or jet mode feedback depends on its Eddington ratio.

\noindent - {\sc Illustris-TNG100}\footnote{\url{https://www.tng-project.org}}. Solves the equations of magneto-hydrodynamics (rather than the hydrodynamics ones as the three previous simulations). Metal radiative cooling also follows \citet{Wiersma09} as in {\sc EAGLE} and {\sc Magneticum}, but the radiation field has contributions from both the background UV and AGN, unlike the other simulations that only consider the background field. The interstellar medium and star formation also follows the model of \citet{Springel03} as in {\sc Magneticum}. Feedback from stars is modelled similarly to {\sc Magneticum}, as kinetic energy injection that is accompanied by decoupling kicked particles from the hydrodynamic calculation (with the main difference being different assumptions for the initial wind velocity). AGN feedback is also modelled as two modes: at high accretion rates black holes inject thermal energy, while at low accretion rates, there is kinetic energy injection, similar to what is done with stellar feedback.

\noindent - TK15 \citep{Taylor15, Taylor17}. This simulation adopts gas cooling, star formation, stellar and AGN feedback prescriptions as well as in the other cosmological simulations, but there is an important difference in the AGN seeding, which results in a different impact on the cosmic star formation rates and stellar populations in galaxies. The main difference with the simulations above is that they included more careful modelling of chemical evolution that considers more sources of chemical pollution. In addition to the standard sources of chemical enrichment included in the other simulations -- namely supernovae core collapse and Type Ia, and asymptotic giant branch star winds -- TK15 also includes hypernovae using yields from \citet{Kobayashi11}.
This simulation covers a relatively small cosmological volume of $(35.7\,\rm Mpc)^3$, which is $7$ to $63$ times smaller than the other simulations presented in Table~\ref{tab:simsdetails}. 
\end{appendix}

\bibliographystyle{pasa-mnras}
\bibliography{biblio}

\end{document}